\documentclass[12pt,preprint]{aastex}
\input epsf.sty

\shorttitle{Spectral features in Ton S180}
\shortauthors{R\' o\. za\' nska et al.}

\begin{document}

\title{The origin of emission and absorption features in Ton S180 
{\it Chandra} observations. }

\author{A. R\' o\. za\' nska and B. Czerny}
\affil{ Nicolaus Copernicus Astronomical Center, Bartycka 18,
            00-716 Warsaw, Poland}
\email{agata@camk.edu.pl; bcz@camk.edu.pl} 

\author{A. Siemiginowska}
\affil{Harvard Smithsonian Center for Astrophysics, 60 Garden
Street, MA02138, Cambridge, USA}
\email{aneta@head-cfa.harvard.edu}
\author{A.-M. Dumont and T. Kawaguchi}
\affil{Observatoire de Paris, Section de Meudon, Place Janssen,
92195 Meudon, France}
\email{Anne-Marie.Dumont@obspm.fr; Toshihiro.Kawaguchi@obspm.fr}

\clearpage

\begin{abstract}

We present new interpretation of Ton S180 spectrum obtained by {\it Chandra}
Spectrometer (Low Energy Transmission Grating).  Several narrow
absorption lines and a few emission disk lines have been successfully
fitted to the data.  We have not found any significant edges
accompanying line emission.  We propose the interpretation of narrow
lines consistent with the paper recently written by Krolik (2002),
where warm absorber is strongly inhomogeneous. Such situation is
possible in so called multi-phase medium, where regions with different
ionization states, densities and temperatures may coexist in thermal
equilibrium under constant pressure.  We illustrate this scenario with
theoretical spectra of radiation transfered through a stratified cloud
with constant pressure (instead of constant density) computed by code
{\sc titan} in plane parallel approximation. Detected spectral features
are faint and their presence do not alter the broad band continuum. 
We model the broad band continuum of Ton S180 assuming an irradiated 
accretion disk with a dissipative warm skin. The set of parameters
appropriate for the data cannot be determined uniquely but models with 
low values of the black hole mass have too hot 
and radially extended warm skin to explain the formation of soft 
X-ray disk lines seen in the data. 

\end{abstract}

\clearpage

\section{Introduction}

Narrow Line Seyfert 1 galaxies (Osterbrock \& Pogge 1985) form a
special class of active galactic nuclei. Narrowness of the optical
lines frequently correlates with the steepness of the soft X-ray
spectra (Boller, Brandt \& Fink 1996) hinting for a connection of this
class with soft state Galactic sources. The overall spectra are
sometimes well described by a disk emission superimposed on a broad
IR-X-ray power law component (Puchnarewicz et al. 2001) but in most
cases the soft X-ray slope seems to be too shallow for an extension of
a disk black body emission. Such a steep soft X-ray spectrum of a
roughly power law shape may form either as a result of Comptonization
(e.g. Czerny \& Elvis 1987) by a moderately hot material (in
comparison with the very hot plasma responsible for the hard X-ray
component in normal Seyfert 1 galaxies) or as a result of reflection
by a strongly ionized material (e.g. Czerny \& \. Zycki 1994, Czerny
\& Dumont 1998). The established presence of warm absorber seen in
many of those objects (Collinge et al. 2001, Kastra et al. 2002, Kaspi
et al. 2002, Kinkhabwala et al. 2002, Leighly et al. 2002, Yaqoob et
al. 2002) complicates the analysis.  However, with the extremely good
spectral resolution of {\it Chandra} and {\it XMM} the issue may be
addressed through the careful analysis of the existing spectral
features.


Ton S180 ($ z = 0.06198$, corresponding to 286 Mpc for Hubble constant 
65 km s$^{-1}$ Mpc$^{-1}$) is an extreme case of a NLS1 with its low 
value of the FWHM of $H_{\beta}$ line ($\sim 900 $ km s$^{-1}$; 
Wisotzki et al. 1995) 
and steep soft X-ray spectrum of photon index $\Gamma=2.68$ 
(Comastri et al. 1998) 
The Galactic absorption toward the object is low
($ N_H = 1.52 \times 10^{20}$ cm$^{-2}$; Stark et al. 1992). The source was
extensively observed in all energy bands. The broad band spectral energy 
distribution (SED) determined by Turner et al. (2002) indicates that most 
of the energy is emitted in
the band $10-100$ eV. Furthermore, {\it HST} data show typical 
emission lines as expected for a Seyfert 1 galaxy, and the lack of 
absorption features. 
High resolution spectra 
obtained by {\it FUSE} reveal UV absorption by O{\sc vi} and the absence 
of neutral hydrogen absorption, indicating  high ionization state for 
the absorbing gas. 

High resolution X-ray spectra of this object were recently obtained
with {\it XMM-Newton} and {\it Chandra} satellites. {\it XMM-Newton}
spectrum was steep and featureless (Vaughan et al. 2002). {\it
Chandra} data also appeared predominantly featureless and well
represented by a single power law with some soft excess (Turner et
al. 2001, Turner et al. 2002).  The shape of the soft excess was
fitted by a {\sc diskbb} model, however plotted residua hinted for the
possible presence of sharp spectral features (Fig. 1 Turner et
al. 2001).

Therefore, in the present paper we reanalyze {\it Chandra} data in
some detail (Sec.~\ref{sec:cha} and \ref{sec:spe}), using a new
response matrix acisleg1D1999-07-22rmfN0002.fits and including any
effects related to the contamination layer on ACIS-S. We note that the
observation of Ton S180 was performed on day 143 of the Chandra
mission when the effects of the contaminant were much smaller than
currently observed. We detected several weak narrow absorption lines
due to the material located at the line of sight to the source.  Also,
several individual disk lines are fitted with equivalent widths in
rough agreement with the predictions of the model of illuminated
accretion disk spectrum presented by R\'o\.za\'nska et al. (2002 a,b).

In Sec.~\ref{sec:apl} we discuss existing models (Kaastra et al. 2002,
Kinkhabwala et al. 2002 and others) for fitting absorption by warm,
partially ionized gas present in many AGN.  We point out,
after Krolik (2001), that the warm absorber may not be the slab of
constant density matter, but it is rather the distribution of regions
with different densities and temperatures.  Such a multi-phase mixture
of gas can exist under constant pressure, as a result of thermal
instabilities (Krolik, McKee \& Tarter 1981).

Till now, the multi-phase models of warm absorber were not
constructed, because popular photoionization codes ({\sc xstar}, 
{\sc cloudy}), compute radiative transfer through the constant density
matter. In Sec.~\ref{sec:mod}, we present new set of models computed 
by the code {\sc titan} (Dumont, Abrassart \& Collin 2000), 
which solves radiative
transfer through the matter in non-LTE approach.  
Introducing assumption of
constant pressure (not density) of warm absorber, we obtain spectra
with many absorption lines from the gas with strong temperature and
density gradient.  Our code is suitable for any optical thickness of
the absorber, nevertheless, we consider regions with total column
density of order of $10^{23} {\rm cm}^{-2}$, as suggested by Krolik
(2001) to be a typical property of warm absorber.

In Sec.~\ref{sec:bbc} we construct two global disk
corona models trying to predict the slope of observed continuum.  The
model with higher black hole mass ($ M=10^8 M_{\odot}$) is more
promising in the sense that emission lines from the disk are expected 
in the spectra, since hot skin does not cover disk completely.

Sec.~\ref{sec:dis} is devoted to the discussion, while   
conclusions are listed in Sec.~\ref{sec:conc}.

\section{{\it Chandra} observations and data reduction}
\label{sec:cha} 

Ton S180 was observed with the {\it Chandra} satellite in December 14,
1999, for about 80 ksec in Low-Energy Transmission Grating in
ACIS-S/LETG configuration, as described by Turner et al. (2001) and
Turner et al. (2002).  The resolution of LETG corresponds to an ${\rm
FWHM} \approx 1200 E_{keV}$ km s$^{-1}$ for the range of energies from
0.1 up to 6 keV.  The source luminosity at this epoch (2--10 keV
unabsorbed flux equal $4.6 \times 10^{-12}$ erg s$^{-1}$ cm$^{-2}$ and
0.3--10 keV unabsorbed flux equal $2.2 \times 10^{-11}$ erg s$^{-1}$
cm$^{-2}$ ) was comparable to that during the {\it Beppo-SAX}
observation (2--10 keV: $4.2 \times 10^{-12}$ erg s$^{-1}$ cm$^{-2}$;
Comastri et al. 1998) and {\it XMM-Newton} observation (0.3 - 10 keV:
$ 2.2 \times 10^{-11} $ erg s$^{-1}$ cm$^{-2}$; Vaughan et al. 2002)
and somewhat fainter than during the 12 days {\it ASCA} monitoring
(2--10 keV: $6.5 \times 10^{-12}$ erg s$^{-1}$ cm$^{-2}$; Romano et
al. 2002).

We obtained the data from the Chandra archive. The data were processed
with the standard CXC data processing pipeline ASCDSVER 6.7.0.  We
analyzed the data using CIAO Version 2.2. and follow Grating Analysis
Thread (CXC web page) to extract the spectrum and create appropriate
effective area files (ARF files).  We combined both sides of the first
order spectrum. We used calibration data base version CALDB~v.2.14,
in particular we applied acisD1997-04-17qeN0003.fits, quantum efficiency
file and acisleg1D1999-07-22rmfN0002.fits as response matrix. This
formally allows us to study the entire energy range between 0.12 till
12.2 keV,
although calibration uncertainties are higher below 0.3~keV, while the
effective area drops significantly above $\sim 4$ keV.  After removing
bad pixels and rebinning the data to achieve at least 20 counts per
channel we
examined the energy between 0.18 and 6~keV.  All spectral fits have
been performed with the XSPEC 9.0 fitting package.

Since the begining of the {\it Chandra} mission ACIS-S detector has
been accumulating a layer of a contaminant which affects any
measurements at low energies (below $\sim 0.8$~keV). Detailed effects
of the contaminant are still being investigated by the Chandra X-ray
Center
\footnote{http://asc.harvard.edu/cal/Acis/Cal\_prods/qeDeg/index.html},
however, it is clear that the overall degradation of ACIS quantum
efficiency and presence of additional absorption features is time
dependent. Ton S180 was observed at the day 143 of the {\it Chandra}
mission and based on the currently available calibration data we
expect that the instrument shows maximum of $\sim$40-15 \%
degradation within 0.28-0.6~keV range (i.e. maximum $\sim 40$ \% in 
0.28~keV and
$\sim 15$ \% in 0.6~keV).  In the following we identify
the possible absorption features due to the absorber in Ton S180 using
the entire available energy range.  We note here that some of the
features at energies close to the edges of the contaminant (C,N,O,F
K-edge) may be affected. We use the available model, {\tt acisabs}, to
include the contamination effects in the analysis.

\section{Spectral analysis}
\label{sec:spe}

\subsection{Continuum}
\label{sec:cont}

In the first step, we fit the  $N_{H}$ value of Galactic absorption 
together with 
a broken power law continuum. Our best fit value of   
$N_{H(Gal.)}=1.83 \pm 0.15 \times 10^{20}$ cm$^{-2}$
is slightly larger than Galactic 
absorption $N_{H(Gal.)}=  1.5 \pm 0.5 \times 10^{20}$ cm$^{-2}$
determined by Dickey \& Lockman  (1990) and used by Comastri et al. (1998) or
than $N_{H(Gal.)}  = 1.52 \times 10^{20}$ cm$^{-2}$
 determined by Stark et al. (1992) 
and used by Turner et al. (1998, 2001).  
All fitted parameters for "plain continuum" are listed in 
Table~\ref{tab0}.

In the next step we fit broken power law with value of Galactic absorption 
fixed at
$N_{H(Gal.)}=1.52 \times 10^{20} $ cm$^{-2}$ 
(see Table~\ref{tab0} second line).
It results in slightly higher $\chi^2$, but the power law index
below 2 keV is  $\Gamma_1=2.69\pm 0.02$, and agrees with the best 
fit  slope $\Gamma = 2.68$ presented by 
Comastri et al. 1998 for Beppo SAX data. Lower value obtained by  
Turner et al. (2001 and 2002), equal to $2.44 \pm 0.04$ resulted from the
use of additional component to model the soft X-ray excess. 

The energy break obtained by us in both models (about $ 1.7 \pm 0.14$ keV) 
is somewhat lower than the value determined by Comastri et al. 1998 
($E_{break}=2.5$), but the spectral slope
above this energy break, $\Gamma_2= 2.26 \pm 0.07 $ agrees with results
 Comastri et al. 1998.  

\begin{figure*}[t]
 \plotone{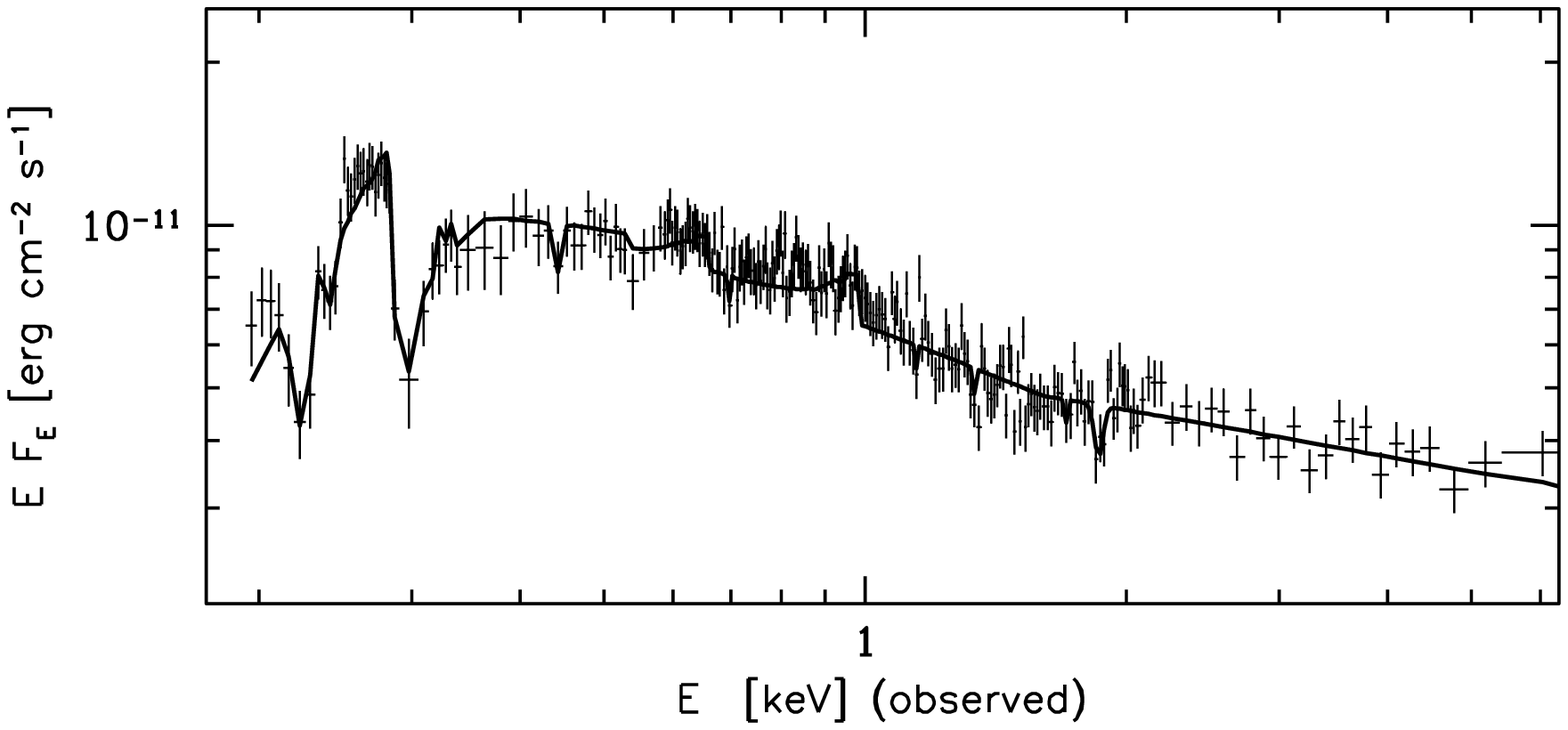} 
 \caption{The LETG spectrum of Ton S180. Solid line denotes
total model after broken power law, Galactic absorption and all lines have
been fitted.}
 \label{fig:cont}
\end{figure*} 

Since both broken power-law indices are very similar to those 
obtained from other satellites we adopt second model (fixed Galactic 
absorption) for further analysis, leaving $\chi^2=1197$ per 1682 d.o.f. 
(reduced $\chi^2=0.7105$).

\begin{deluxetable}{llllll}
\renewcommand{\arraystretch}{1.5}
\tablenum{1}
\tablewidth{155 mm}
\tablecaption{LETG continuum fit in the 0.1-- 6. keV energy range
\label{tab0}}
\tablehead{
\colhead{Model} &
\colhead{$N_{H(Gal.)}$ $10^{20}$ cm$^{-2}$} &
\colhead{$\Gamma_1$} &
\colhead{$E_{break}$ keV} &
\colhead{$\Gamma_2$} &
\colhead{$\chi^2$/d.o.f.}  
}
\startdata
Plain Cont. &$ 1.8277^{+0.152}_{-0.149}$ & $ 2.755^{+0.038}_{-0.0367}$ & 
$ 1.659^{+0.153}_{-0.109}$& $2.265^{+0.062}_{-0.065}$ & 1185/1681\\
Fixed $N_{H(Gal.)}$  & 1.52  & $ 2.689^{+0.022}_{-0.019}$ & 
$ 1.771^{+0.134}_{-0.169}$ & $2.261^{+0.074}_{-0.070}$& 1197/1682 \\ 
With lines & $ 1.527^{+0.1512}_{-0.14}$ &  $2.682^{+0.039}_{-0.0359}$ & 
$ 1.604^{+0.12}_{-0.147}$ & $2.295^{+0.0549}_{-0.0564}$ & 932/1624 \\

\enddata
\end{deluxetable}

After fitting continuum, there is still
some structure in the residuals of the soft X-ray spectrum, suggesting the presence of
absorbing matter along the line of sight toward the source. 
Therefore, we have tried to determine column density of this gas
in the simplest way, by fitting O{\sc vii}, O{\sc viii}, and C{\sc vi}
edges often seen in Seyfert galaxies (Mrk 509 Yaqoob et al. 
2002, NGC 3783 Kaspi et al. 2002). 
Surprisingly, such edges appeared to be very weak in the data of Ton S180.
We were able only to give upper limits for 
$\tau_{\rm O{\sc vii}}$ 
$\tau_{\rm O{\sc viii}}$, and 
$\tau_{\rm C{\sc vi}}$ of order of $10^{-4}$, which 
practically indicates the lack of those edges. 

\subsection{Spectral features}

\subsubsection{Application of standard warm absorber}
\label{sec:apl}

X-ray absorption lines appear to be common feature in  many AGN.
The widely acceptable explanation is the 
existence of warm, ionized gas which is illuminated by
the energetic radiation from nucleus.
Trying to find properties of warm absorber we need proper tools to 
fit those data.
Till now, models of absorption by the 
partially ionized plasma available in XSPEC ({\sc absori}, {\sc photo},) 
do assume slab with constant density. 

In  {\sc absori}, the radiative transfer of lines is not computed at all, 
only photoionization
cross sections are determined and absorption ``on the spot'' 
is computed $F_{out}=F_{in} exp(-\tau_{line})$ (Done et al. 1992).
Constructed by Kinkhabawa et al. (2002) local {\sc photo} model 
(http://xmm.\-astro.\-columbia.\-edu/research.html) neglects 
photo-electric absorption, but allows for individual line absorption,
treating each ionic column density separately with an initially 
unabsorbed power law. 
This method produces consistent results only when absorbing 
matter is optically thin, and transmitted spectra do not
display photo-electric edges. 
Furthermore, they have assumed that plasma is optically
thin to reemitted photons.

In the aim to find evidences for warm absorber in Ton S180, we have applied 
{\sc absori} to our  data.
Allowing for three parameters to be fitted in this model, $\chi^2$ 
decreases by 38 (now reduced $\chi^2=0.747$).
The fit indicates that the medium has column density
 $7.34^{+0.23}_{-0.22} \times 
10^{21}$ cm$^{-2}$
and ionization parameter ($\xi =4178^{+2082}_{-1328}$) 
for the fixed default temperature $T_{abs}=3 \times 10^4$ K.  
This model, however, leaves considerable residual trends and does not allow
for a deeper insight into the ionization state of the absorbing material.
It is not surprising since the model is based on simple 
computations of photoionization cross sections without accurate
transfer through the medium.

We have applied {\sc photo} to our analysis. 
The model again indicates very low 
column densities of each ion, about $10^{14}$ cm$^{-2}$ for C{\sc vi}, 
O{\sc vii}, and O{\sc viii}.
We conclude that most probably, the warm absorber in Ton S180
is more complicated than optically thin constant density slab.
It may be the stratified density gas, where accurate radiative transfer, 
including all photoionization
processes, should be calculate to model transmitted spectrum.
Such models are not available for fitting the data yet, so two or three
separate  absorbers with different constant densities and ionization states
are used together to explain spectral features
in some AGN  (Kaastra et al. 2002 , Kaspi et al. 2002).
This motivates us to present in Sec.~\ref{sec:mod} the model of the 
multi-phase gas with stratified density and temperature, where 
the radiative transfer of incident power law is carefully computed. 
The model is not
ready for fitting yet, but gives interesting results based on equivalent widths of 
various lines. Observed line properties in the 
Ton S180 data are determined in the section below.

More advanced modeling was done 
with the code {\sc xstar} (developed by Kallman \& Krolik 1999) 
which allows to 
estimate the total column densities of absorbed gas.
Models are computed assuming constant density slab. 
Such models are very useful for reconstruction of the flat part of 
curve of growth (Yaqoob et al. 2002), and for subsequent estimation
of turbulent velocities of the absorber.
Nevertheless, at the last stage of analysis separate Gaussian components 
are fitted to the data.

\subsubsection{Absorption lines}
\label{sec:abs}

\begin{figure}
 \plotone{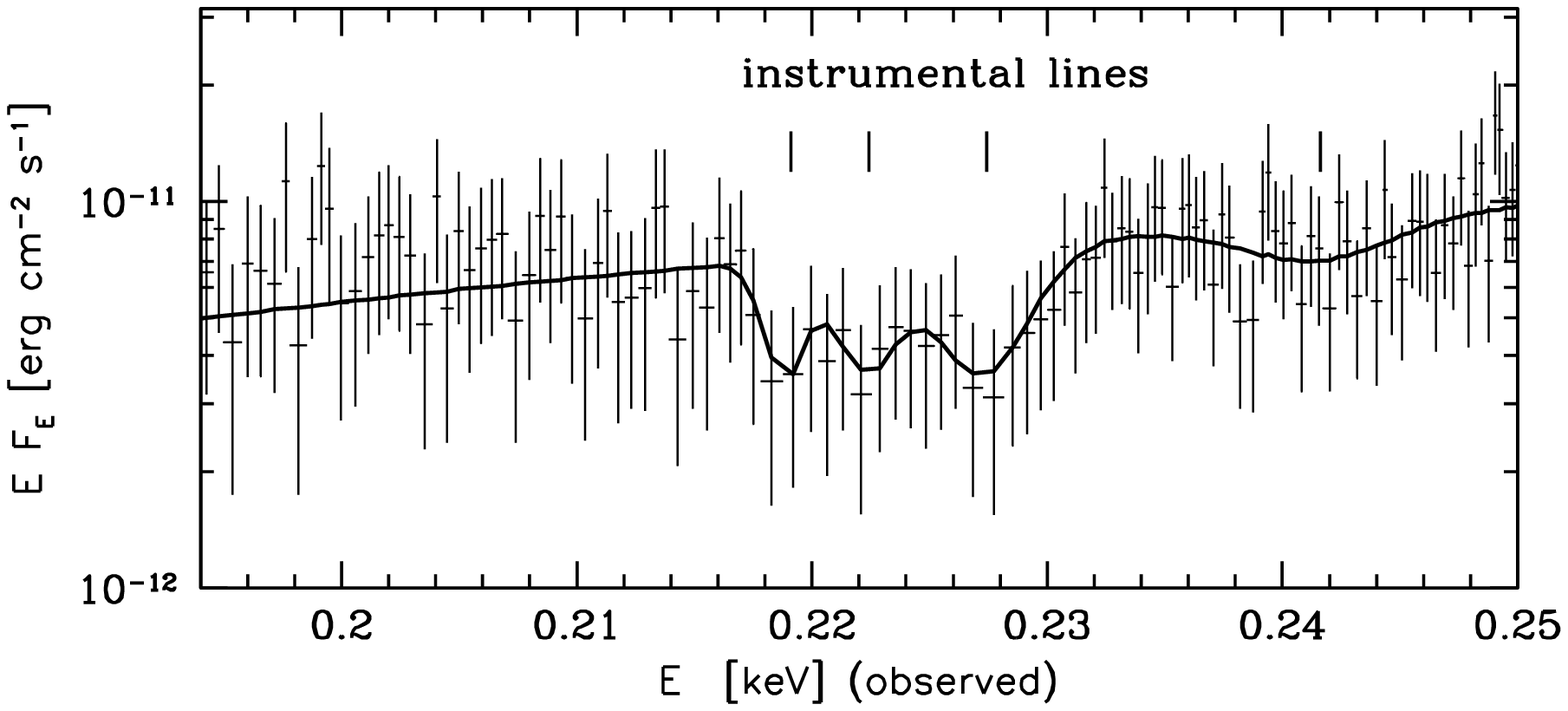} 
 \caption{Observed spectrum with the best fit model for 0.19-0.25 keV}
 \label{fig2:cont}
\end{figure} 
\begin{figure}
  \plotone{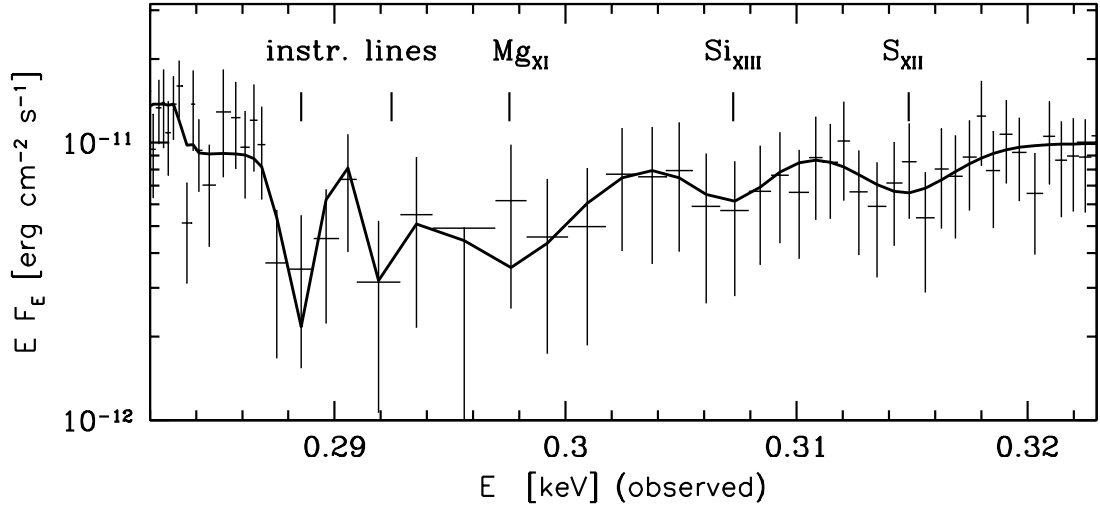}
  \caption{As Fig.~\ref{fig2:cont}, but for 0.282-0.32 keV}
 \label{fig3:cont}
\end{figure}
\begin{figure}
 \plotone{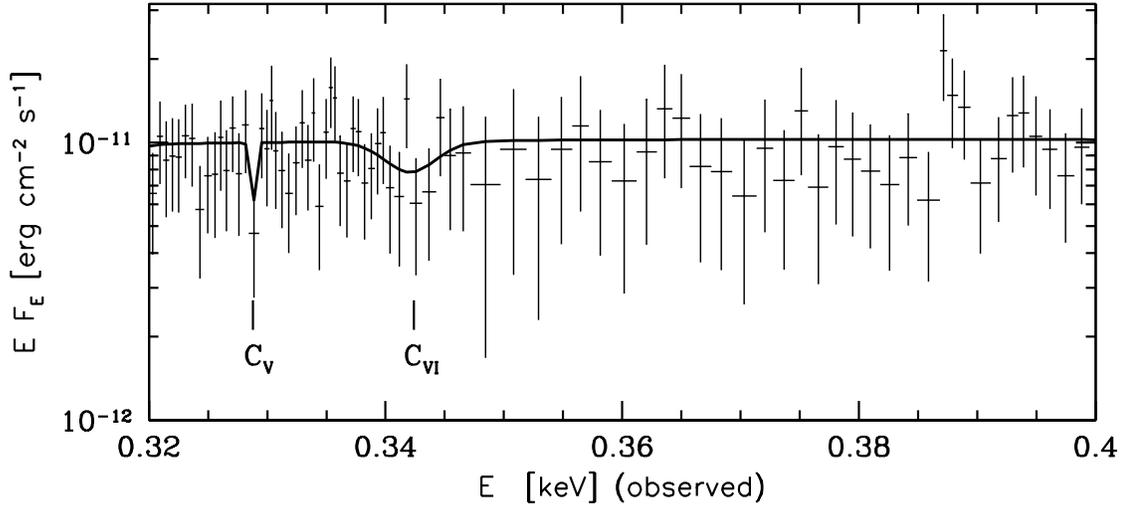}
  \caption{As Fig.~\ref{fig2:cont}, but for 0.32-0.4 keV}
 \label{fig4:cont}
\end{figure}
\begin{figure}
 \plotone{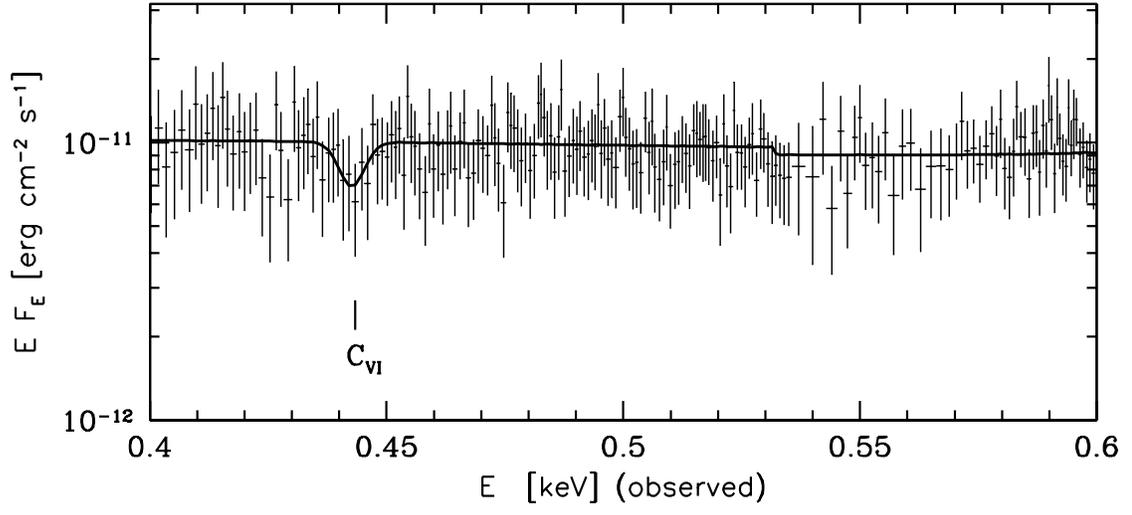}
  \caption{As Fig.~\ref{fig2:cont}, but for 0.4-0.6keV}
 \label{fig5:cont}
\end{figure}
\begin{figure}
   \plotone{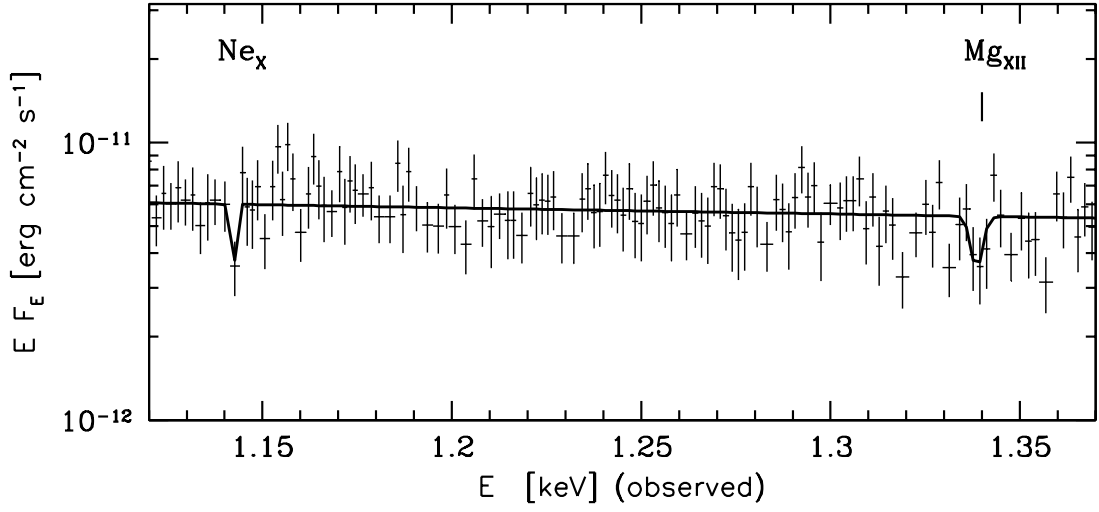}
  \caption{As Fig.~\ref{fig2:cont}, but for 1.12-1.37 keV}
 \label{fig6:cont}
\end{figure} 
\begin{figure}
   \plotone{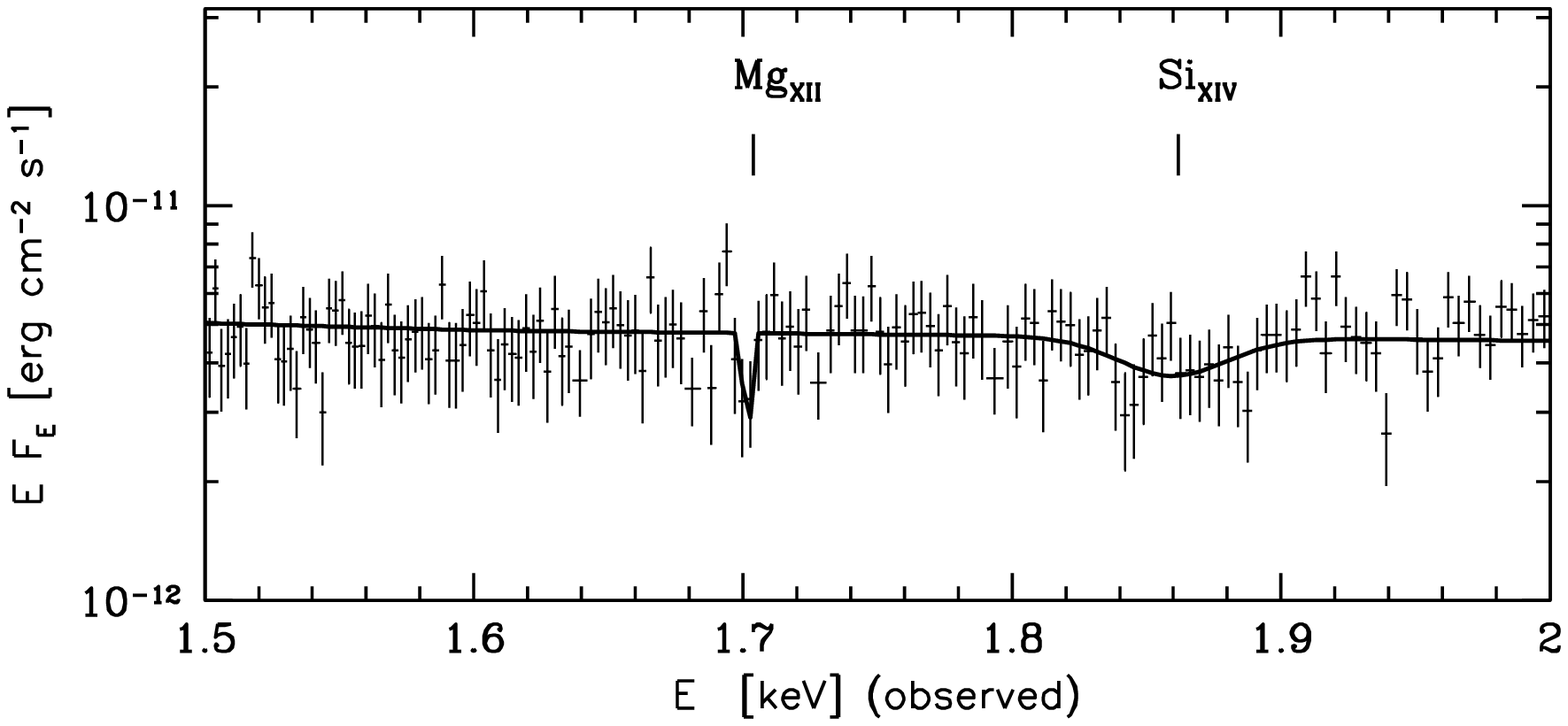}
   \caption{As Fig.~\ref{fig2:cont}, but for 1.5-2 keV}
 \label{fig7:cont}
\end{figure} 

We have successfully fitted several absorption lines using standard Gaussian 
model for each line separately (see Figs.~\ref{fig:cont}, \ref{fig2:cont}, 
\ref{fig3:cont}, \ref{fig4:cont}, \ref{fig5:cont}, \ref{fig6:cont}, and 
\ref{fig7:cont}. {\bf Gaussian fit requires three free parameters, 
energy of the
line centroid, Gaussian width and normalizations, which has to be 
negative in case of absorption lines.} We include in the fit only
those features which are significant.
All absorption lines are presented in Table~\ref{tab1}. 
For each line we list its central energy in keVs, equivalent width (EW) in eVs,
the decrease of $\chi^2$, 
best identification with observed and rest frame energy, and with possible 
blends.

\begin{figure*}
  \plottwo{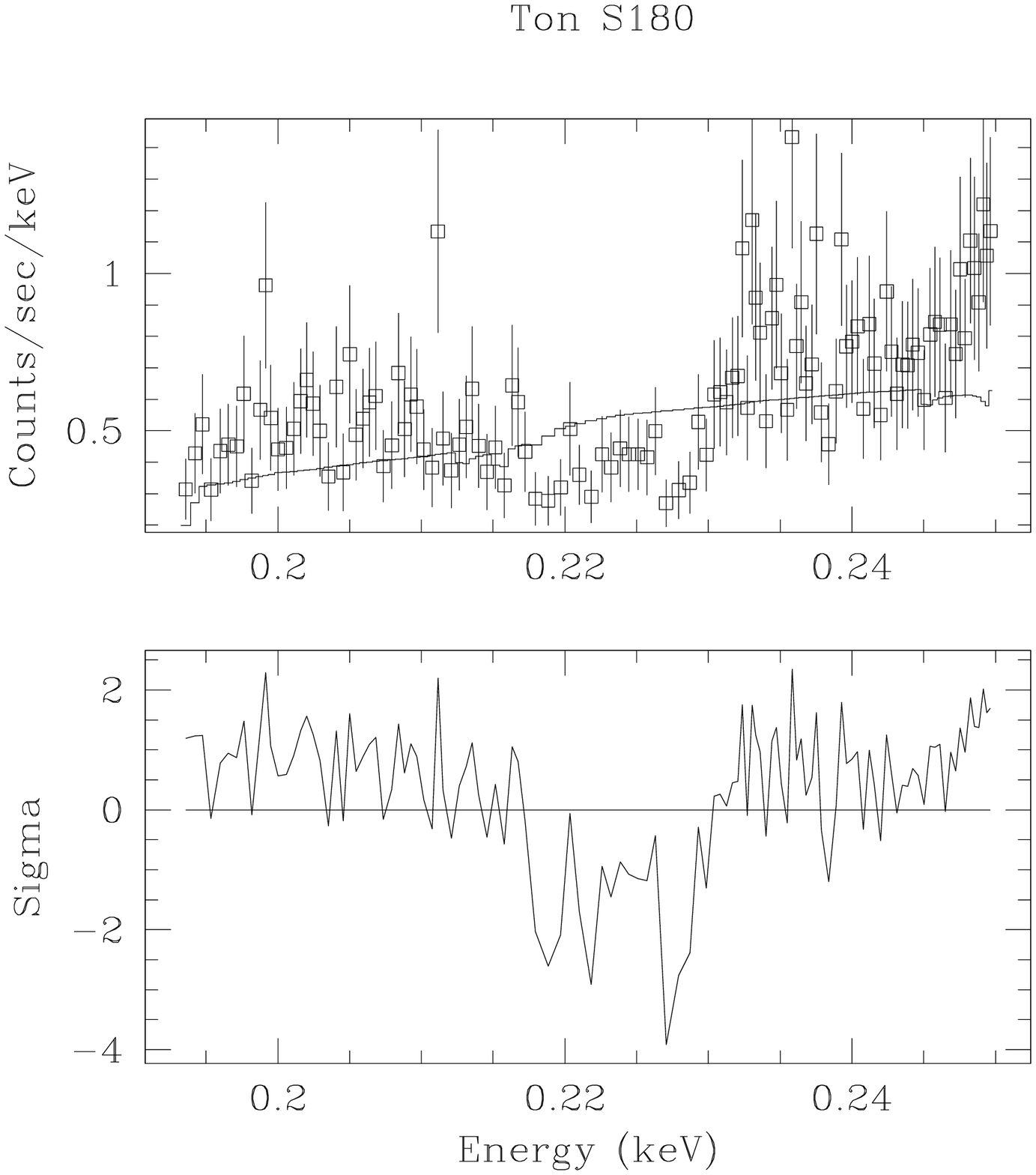}{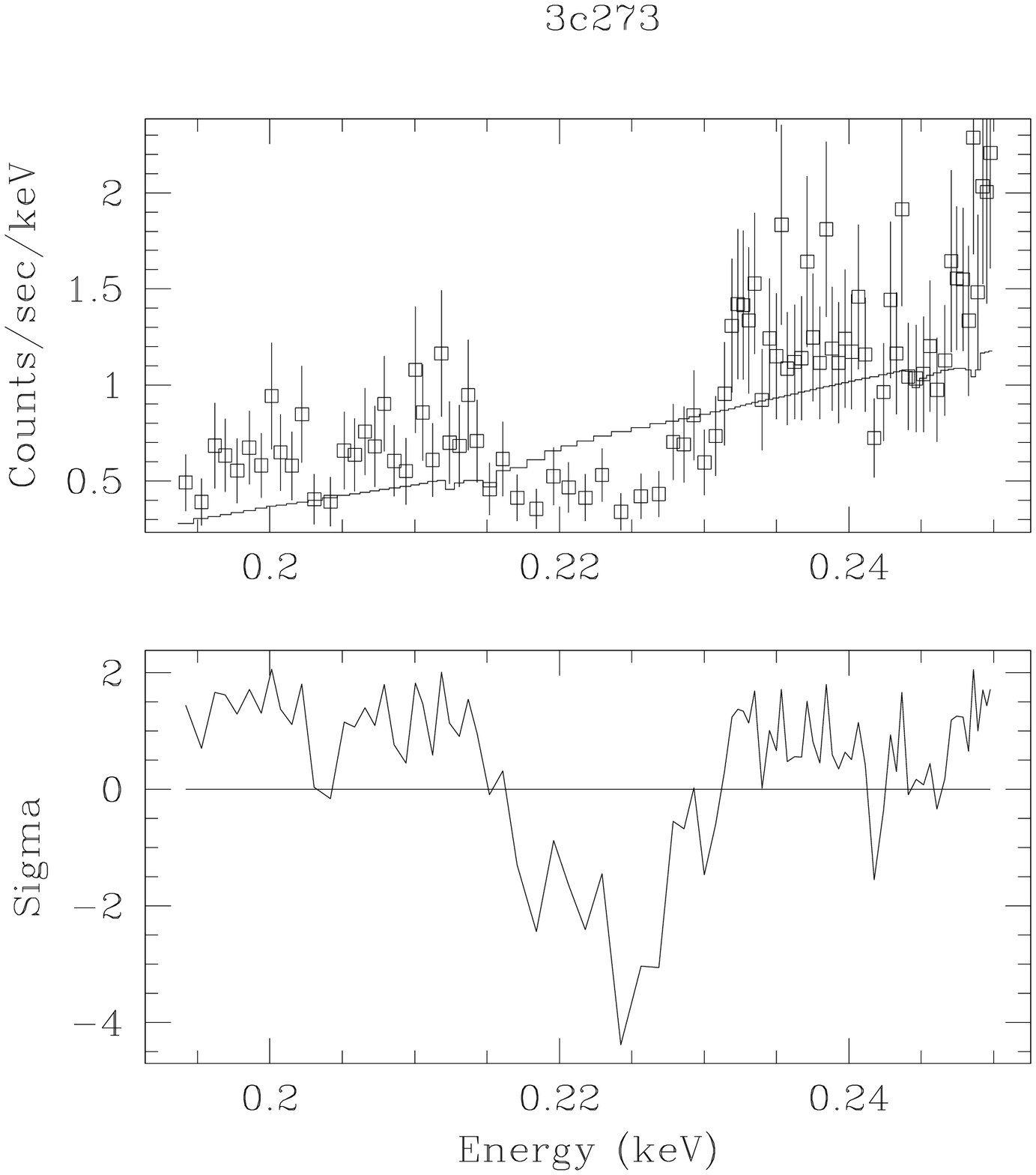}
  \caption{The comparison of {\it Chandra} data of Ton S180 to the 
       {\it Chandra} data of 3c273 for energies between 0.20 to 0.24 keVs. 
    Similar structure in residuals around 0.22 keV is present in both objects.}
 \label{por:22}
\end{figure*}

Some of those lines, especially the softest ones, may be caused by
dust presented on Chandra mirrors.  This should show up a common
feature in all objects even those with different redshifts. To check
whether lines are real or instrumental, we compare observations of
Ton S180 with {\it Chandra} data of quasar 3c273 observed two weeks
latter with the same configuration as Ton S180.  
Exemplary residuals (for soft
range between 0.2-0.24 keV) for both objects are shown in
Fig.~\ref{por:22}. In both cases local power-law was fitted with 
fixed value of Galactic absorption appropriate for each object. 
 The same features around energy
0.22 keV are present as well in 3c273 with redshift 0.156 as in
Ton S180. 
We repeated such analysis in the whole energy range. All
possible instrumental lines are labeled in Table~\ref{tab1}.

\begin{figure*}
 \plottwo{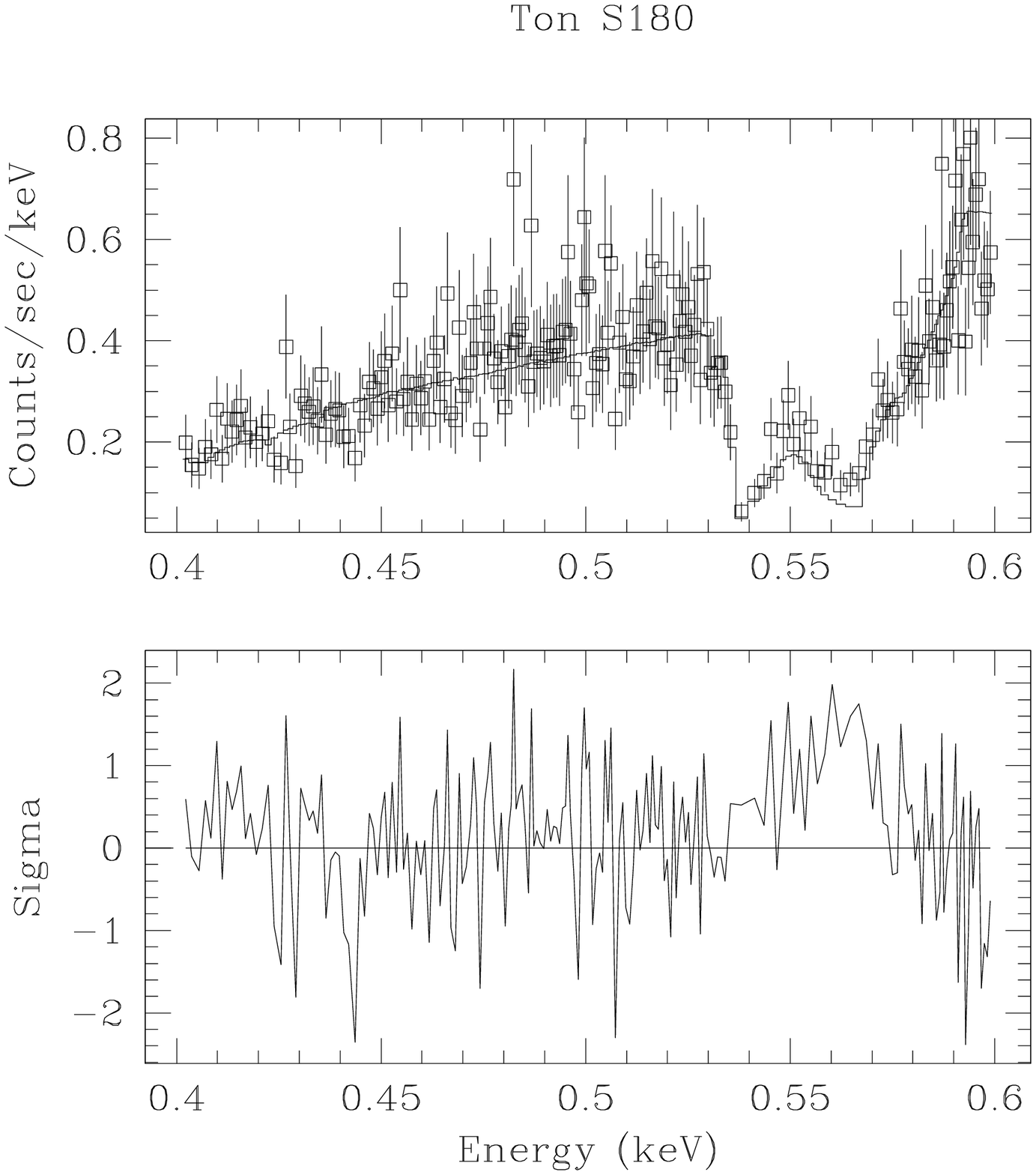}{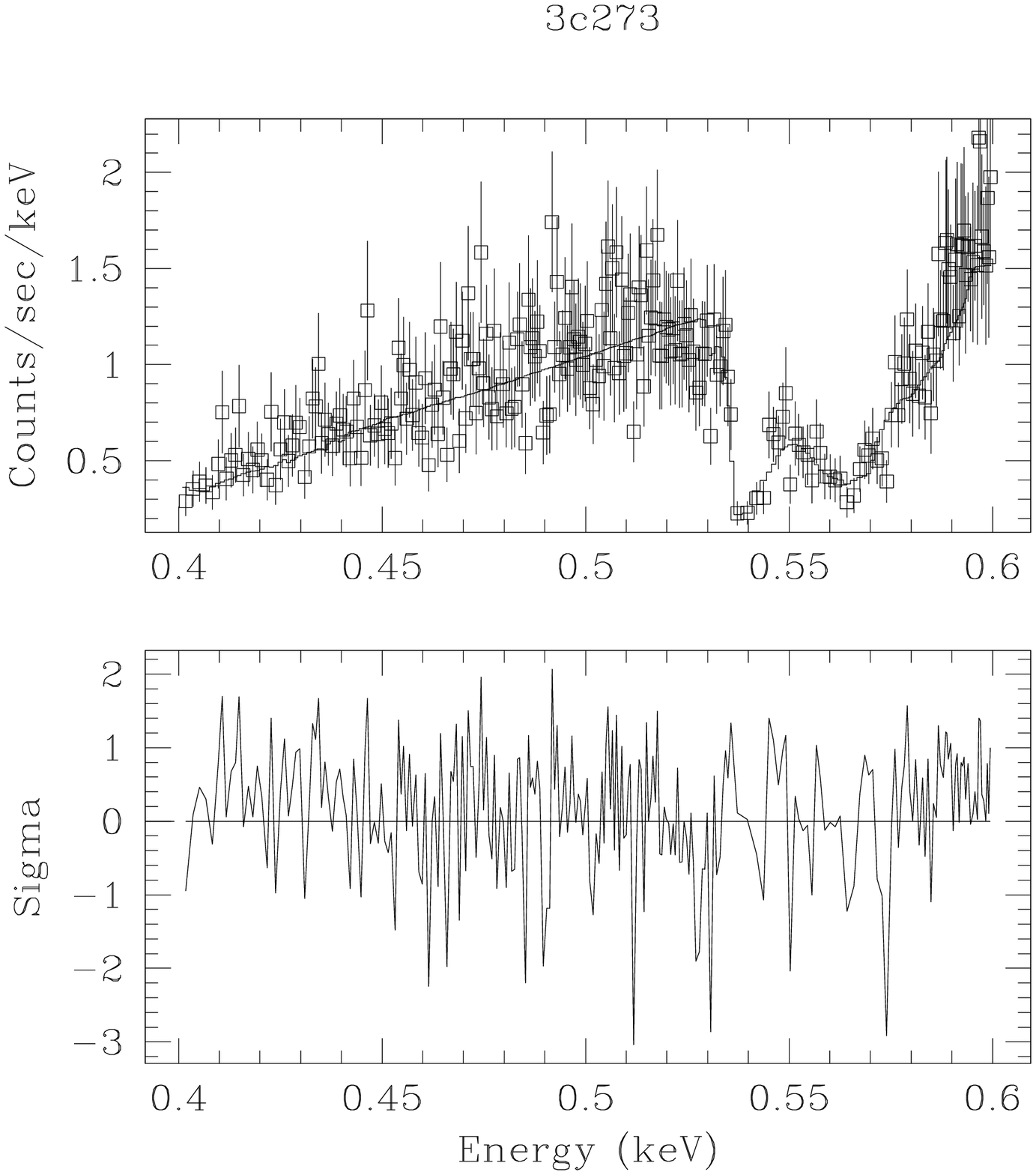}
  \caption{The comparison of {\it Chandra} data of Ton S180 to the 
  {\it Chandra}  data of 3c273 for energies between 0.4 to 0.6 keVs.
    Two sigma structure presented in Ton S180 around 0.44 keV 
  is clearly absent in 3c273 data.}
 \label{por:44}
\end{figure*}

Some of the detected features are seen only in Ton S180 spectrum and
therefore seem to be real, intrinsic to the source. 
For example, in Fig~\ref{por:44} 
we show that line C{\sc vi} with observed energy 0.4428 keV which is
detected in Ton S180 data is not seen in 3c273.  

Lines are very faint. They usually have equivalent width less or about
2 eVs, and several of them are just marginal detections, as clearly
seen from the line EW error, given at 90\% confidence level
({\bf each error is computed for the one parameter of interest}).  Only
two of them are stronger, but from theoretical predictions 
by analysing our modeled spectra from Sec~\ref{sec:mod}, they are not 
blends. All possible blends according of our
modeled specta obtained with spectral resolution 100 are listed in 
Table~\ref{tab1}. 
There is therefore no major
discrepancy between the {\it Chandra} results and the {\it XMM}
observations (Vaughan et al. 2002) where only upper limits of order of
1 eV at energies 0.35 - 0.7 keV were provided. Most lines originate
from hydrogen or helium like ions indicating that matter is highly
ionized. There is an absence of lines from mildly ionized spices.

Interestingly, we found two different transitions from the
same element at the given ionization level, 
(C{\sc vi}$_{(1-2)}$ and  
C{\sc vi}$_{(1-5)}$ 
where i -- j denotes number of levels from and to which the 
transition proceeds).
This is helpful in the aim to estimate the optical thickness
of C{\sc vi} ions and then total optical thickness of the warm absorber. 
  
If the medium is optically thin in those lines (i.e. weak line case),  we
are on the linear part of the curve of growth, 
where equivalent width
is proportional to the column density $N_i$ of the absorbed ion:
\begin{equation}
EW_{\nu}=\frac{\pi e^2}{m_e c} N_i f_{i j},
\label{equ:column}
\end{equation} 
where $f_{i j}$ is the oscillator strength transition and stimulated emission 
has been set equal to unity.
Thus, for the given population of ions at the same ionization level,
the ratio of equivalent widths is given by:
\begin{equation}
\frac{EW_{i j}}{ EW_{i k}} = \frac {f_{i j}} {f_{i k} }.
\end{equation}  
Since the oscillator strengths are strictly atomic features independent
from the radiation field and are calculated theoretically
we use them to find the ratio
\begin{equation}
\frac{EW[{\rm C VI}_{(1-2)}]}{EW[{\rm C VI}_{(1-5)}]} =  29.86. 
\label{equ:tra}
\end{equation}
On the other hand, the same ratio determined for Ton S180 
(see Table~\ref{tab1})
equals $ 0.62^{+3.53}_{-0.59} $. 
Error is computed assuming most pessimistic 
case, when numerator has its maximal value $EW+\Delta EW$, and denominator has 
its minimal value $EW-\Delta EW$. Although the error is huge, the value of 
the ratio determined from observation is still far from theoretical predictions
for optically thin medium (Eq.~\ref{equ:tra}).
Our theoretical spectra (see Sec.~\ref{sec:mod}) suggest that  
C{\sc}$_{(1-2)}$ 
line is possibly a blend with Mg{\sc xii}, Si{\sc xiv} and most strongly with C{\sc v}.
We have computed that real EW of the C{\sc}$_{(1-2)}$  line can be even
up to four times lower than the EW of the whole blend. 
But this fact even lowers the observed ratio of EWs of two carbon lines.  
 
The straight conclusion is that the emitting region is optically thick for
those lines and
the transfer in lines should be taken into account for 
proper modeling of the warm absorber.
It is also possible that the identification of the
lines is not correct or the line is a blend of the carbon line with
another transition. Nevertheless, we further explore the possibility that
the carbon line ratio is determined correctly.

Therefore, the Equation~\ref{equ:column} gives only a lower limit for the 
ion column density, equal to $ 3.6 \times 10^{16} $ cm$^{-2}$ 
from the C{\sc vi}$_{(1-2)}$ line and $ 1.7 \times 10^{18} $ cm$^{-2}$  from  
C{\sc vi}$_{(1-5)}$, which under assumption of all carbon being in the form 
of C{\sc vi} ions and with solar abundances translates into the lower limit 
for the warm absorber column equal $ 1.1 \times 10^{20} $ cm$^{-2}$ and
$ 5.1 \times 10^{21} $ cm$^{-2}$, correspondingly. 


 
Exact determination of the column density is complicated.  At the flat
curve of growth we need additional information about velocity
dispersion $b$ to indicate column densities.  In the case of
considered data this is very difficult, because $1 \sigma$ error of
full width of half maximum (FWHM) is too high to estimate $b$
parameter.  For instance for S{\sc xii} line, presented in
Fig.~\ref{fig3:cont}, $FWHM = 5^{+16}_{-1} \times 10^3$ km/s.  This fact also
precludes us to integrate over fitted line profile in the aim of
determining exact flux emitted in this line.  So, we are unable to
compute exact optical depth of the line and therefore the column
density of absorbing ion.  The number of lines is to small to fit full
model (for instance {\sc xstar}).

Another lower limit on column density of C{\sc vi} ion can be derived knowing 
that the optical thickness of the C{\sc vi} lines is higher than 1.
In such a case we can write:
\begin{equation}
N_i > \frac{m_e c}{ \pi e^2 f_{i j}} FWHM = 
\frac { 1.665 m_e \nu_0}{\pi e^2 f_{i j} } \sqrt{ \frac {2 k T }{ A_i m_H}},
\end{equation}
where $T$ is the temperature of the medium and $A_i$ is atomic mass of the ion.
For the C{\sc vi}$_{(1-5)}$ line we obtain:
\begin{equation}
N_{CVI} > 6.33 \times 10^{13} \sqrt{T}.   
\end{equation}  
For the temperature approximately equal to $10^6$ K this value is twice higher 
than the
one determined by Mason et al.(2002) in case of Mrk 766. This limit is over an
order of magnitude weaker
than the limit obtained directly from Equation~\ref{equ:column}. FWHM 
estimated from the $T \sim 10^6$ K is quite narrow and the line width is 
rather determined by much larger turbulent motion.

The value of reduced $\chi^2$ was equal 0.6363 after fitting all possible
absorption lines.  

\subsubsection{Disk emission lines}


\begin{figure}
   \plotone{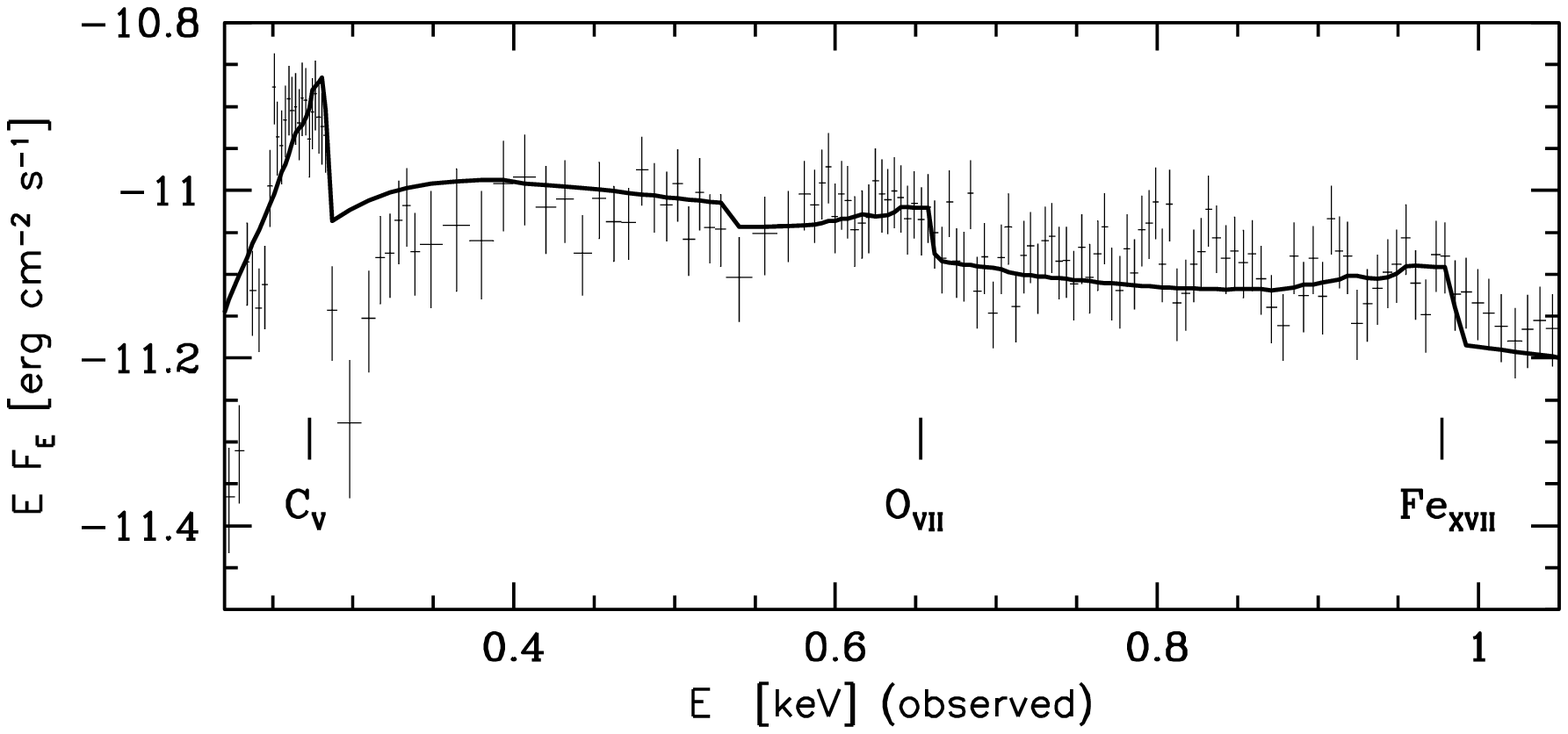}
   \caption{As Fig.~\ref{fig2:cont}, but for 0.22-1.05 keV. Three emission
lines fitted to the spectrum of Ton S180 are seen.}
 \label{fig8:cont}
\end{figure}

After absorption lines have been identified, we were able to fit a few
emission lines from an accretion disk. We applied the standard {\sc
diskline} model with fixed power law index in the radial dependence of
the irradiating flux $\beta=2.7$. We assumed $r_{in} = 6 R_{g}$ and
$r_{out}= 1000 R_{g} $ (where $R_{g}= GM/c^2$). We have found the best
value of inclination angle $i=30^{o}$ for the line C{\sc v}, and we
adopted it (keeping frozen) for the other lines.  Results are listed
in Table~\ref{tab2}, and shown in Figs.~\ref{fig:cont}, and \ref{fig8:cont}.

The last column of Table~\ref{tab2} presents theoretical values
of EWs of the same emission lines computed for accretion 
disk illuminated by hard X-ray flux equaled to the local disk flux 
(full model presented in R\'o\.za\'nska et al.  2002 a,b).
The model takes into account harder incident radiation ($\Gamma=1.9$),
than it is seen in case of Ton S180, and solar metal abundances.
For moderate illumination in the model we have  found for instance 
the equivalent  width of Ne{\sc x} equal 11 eV, 
O{\sc viii} equal 18.2, Mg{\sc xii} equal 10.5,
and Fe{\sc xvii} about 8 eV, or many others at the same order of magnitude
in respect to the reflected continuum (factor of 2-3 smaller in respect to the
total continuum). Models do not predict lines with EWs higher 
than $\sim 20$ eVs, when the disk is in the hydrostatic equilibrium
(R\'o\.za\'nska et al.  2002 a,b; see also figures in  Nayakshin, Kazanas
\& Kallman 2000,
Ballantyne, Ross \& Fabian 2001). 
Higher values 
can be only obtained from constant density medium, as discussed by 
Ballantyne, Ross \& Fabian (2002), particularly when adopting enhanced
metalicity. Emission lines also become by a factor of a few stronger when the
dynamical flare model is adopted, i.e. the time-scale of the flare is shorter 
than the time-scale for a disk surface layer to achieve a hydrostatic 
equilibrium (Collin et al. 2002). However, in that case lines should be narrow
and not represented well by a {\sc diskline} shape.

Soft disk line detection is therefore a promising tool to 
constrain the conditions within the cold reprocessing medium. Detection of 
standard broad disk lines presented in this paper, if real, suggest no 
strong departure from hydrostatic equilibrium and the extension of the cold
disk roughly down to the marginally stable orbit.

After all lines have been fitted, we fit again broken power-law with
Galactic absorption. 
Results presented in Table~\ref{tab0} last verse indicate that lines do not
affect continuum very strongly. 
The final  value of reduced $\chi^2$ appeared to be
0.5531.

\begin{deluxetable}{llcc}
\renewcommand{\arraystretch}{1.5}
\tablenum{2}
\tablewidth{132 mm}
\tablecaption{Absorption lines fitted to the {\it Chandra} data of Ton S180
\label{tab1}}
\tablehead{
\colhead{Observed energy (keV)} &
\colhead{EW (eV) } &
\colhead{$\Delta \chi^2$  } &
\colhead{Best identification $^{(*)}$}  
}
\startdata
$0.2188^{+0.005}_{-0.014}$  & $  1.39 \pm 0.87  $  & 7.42  & {\it instumental}   \\
\hline
$0.2222^{+0.0011}_{-0.0010}$ &  $ 2.05 \pm 1.38 $  & 6.6 & {\it instumental}   \\
\hline 
$0.2272^{+0.0008}_{-0.0009}$ & $ 3.62 \pm 1.29 $ & 22.29 & {\it instumental}   \\
\hline 
$0.2414^{+0.0028}_{-0.0036}$  & $ 1.87 \pm 1.08 $  & 6.34 & {\it instumental}  \\
\hline 
$0.2884^{+0.0007}_{-0.0006} $  & $ 1.82 \pm 0.83 $  & 15.35 & {\it instumental}  \\ 
\hline 
$0.2923^{+0.0012}_{-0.0012}$ & $ 1.39 \pm 1.17 $  & 6.92 &{\it instumental}  \\
\hline 
$0.2974^{+0.0018}_{-0.0022}$ & $ 6.31 \pm 3.70 $ & 14. & Mg{\sc xi} (0.2971; 0.3155)\\
\hline 
$0.3071^{+0.0026}_{-0.0030}$& $ 2.26 \pm 1.95$ & 5.18 & Si{\sc xiii} (0.30878; 0.3279)\\
\hline 
$0.3147^{+0.018}_{-0.030}$ & $ 2.05 \pm 1.28 $ & 7.55  &S{\sc xii} (0.31899; 0.3387) \\
\hline 
$0.3285^{+0.0009}_{-0.0006}$ & $ 0.39 \pm 0.50 $ &3.47 &C{\sc v} (0.33386; 0.3545)  \\ 
\hline         
$0.3421^{+0.0037}_{-0.0045}$ & $1.38 \pm 1.30 $ & 4.45 & C{\sc vi} (0.3459; 0.3673) \\
 & & &  blend with Mg{\sc xii} (0.368), \\
 & & & Si{\sc  xiv} (0.3703), C{\sc v} (0.3709) \\
\hline 
$0.4426^{+0.024}_{-0.024}$  & $  2.22 \pm 1.58 $ & 5.16 & C{\sc vi} (0.4428; 0.4702) \\
\hline                                
$0.6963^{+0.015}_{-0.032}$ & $ 0.74 \pm 0.69 $ & 3.47 & Fe{\sc xvii} (0.6868; 0.72932) \\
 & & &  blend with S{\sc xvi} (0.731) \\
\hline 
$1.142^{+0.0034}_{-0.0095}$ & $ 0.992 \pm 1.18 $ & 3.29 & Ne{\sc x} (1.1387; 1.209) \\
 & & &  blend with Fe{\sc xxv} (1.206), \\
 & & &  Fe{\sc xxv} (1.208), and Fe{\sc xxv} (1.213) \\
\hline 
$1.339^{+0.0030}_{-0.0024}$ & $ 1.48 \pm 1.84 $ & 2.95 & Mg{\sc xii} (1.3861; 1.472) \\
 \hline 
$1.702^{+0.030}_{-0.032}$ & $ 1.88 \pm 2.38 $  & 3.7 & Mg{\sc xii} (1.7327; 1.840) \\
 & & &  blend with Si{\sc xiii} (1.853) \\
\hline                                       
$1.860^{+0.013}_{-0.016}$ & $ 10.6 \pm 5.24 $ & 6.09 & Si{\sc xiv}  (1.8831; 2.000) \\
\enddata

\tablenotetext{(*)}{
Expected observed energy and the rest frame energy in keVs for 
identifined lines are
given in parenthesis, for blends only rest frame energy is listed.}

\end{deluxetable}

\begin{deluxetable}{lllcc}
\renewcommand{\arraystretch}{1.5}
\tablenum{3}
\tablewidth{155 mm}
\tablecaption{Emission disk lines fitted to the {\it Chandra} data of Ton S180
\label{tab2}}
\tablehead{
\colhead{Observed energy (keV)} &
\colhead{EW (eV) } &
\colhead{$\Delta \chi^2$} &
\colhead{Best identification} &
\colhead{EWs from illuminated disk{*}} 
}
\startdata

$0.2704^{+0.002}_{-0.0007} $ & $ 15.3 \pm 2.63 $ & 72.73  &  C{\sc v} (0.2899; 0.3079)  &  8.69 \\ 

$0.6300^{+0.014}_{-0.01}$  & $  13.5 \pm 4.89 $ & 21.26 & O{\sc viii} (0.6149; 0.653) & 19.0 \\

$0.9403^{+0.007}_{-0.021}$ &  $30.7 \pm 7.41 $ & 46.26 & Ne{\sc x} (0.961; 1.020) & 11.29\\


\tablenotetext{*}{
Theoretical EWs are computed in respect to reflected continuum}

\enddata
\end{deluxetable}

\section{Proposed Model of the Warm Absorber}
\label{sec:mod}

\begin{figure}[t]
  \plotone{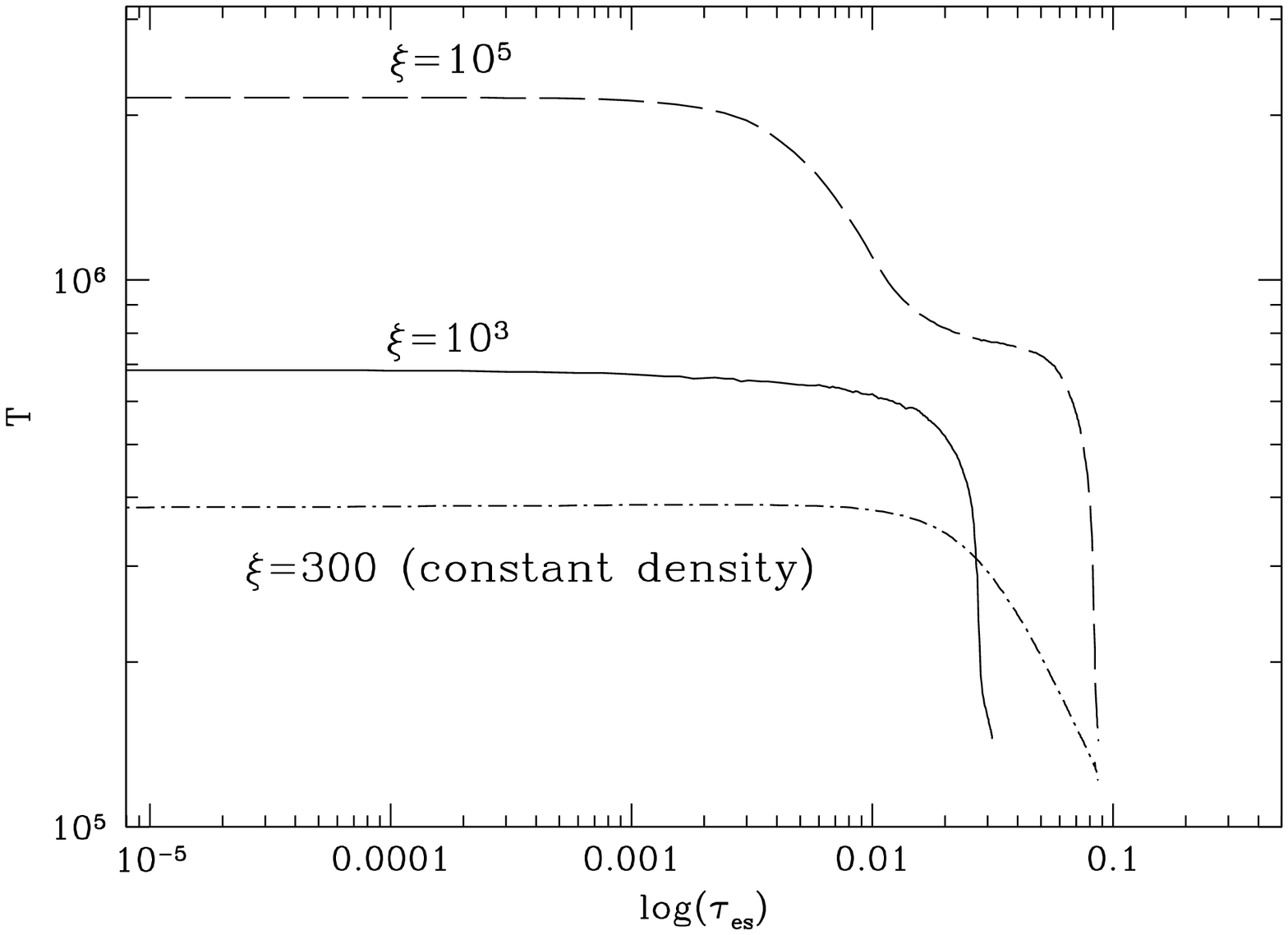}
  \caption{The temperature distribution for models  with  $\xi=10^5$ 
    and $N_{tot} =1.7 \times 10^{23}{\rm cm}^{-2} $ (dashed line),  
     $\xi=10^3$ and $N_{tot} =6.8 \times 10^{22}{\rm cm}^{-2} $ (solid line).
 Constant density case for $\xi=300$ and $N_{tot} =1.7 \times 10^{23}{\rm cm}^{-2}$ is
 presented by dotted-dashed line.}
 \label{fig:mod}
\end{figure} 

In this section, we present models of warm absorber, allowing density,
temperature and so consequently ionization state to be stratified 
(Krolik 2002). The only assumption which we made 
is the constant pressure. Such an approach is more realistic
than the constant density model,
since thermal instabilities can produce gas with steep 
ionization structure keeping all zones in pressure equilibrium.

\begin{figure}[t]
 \plotone{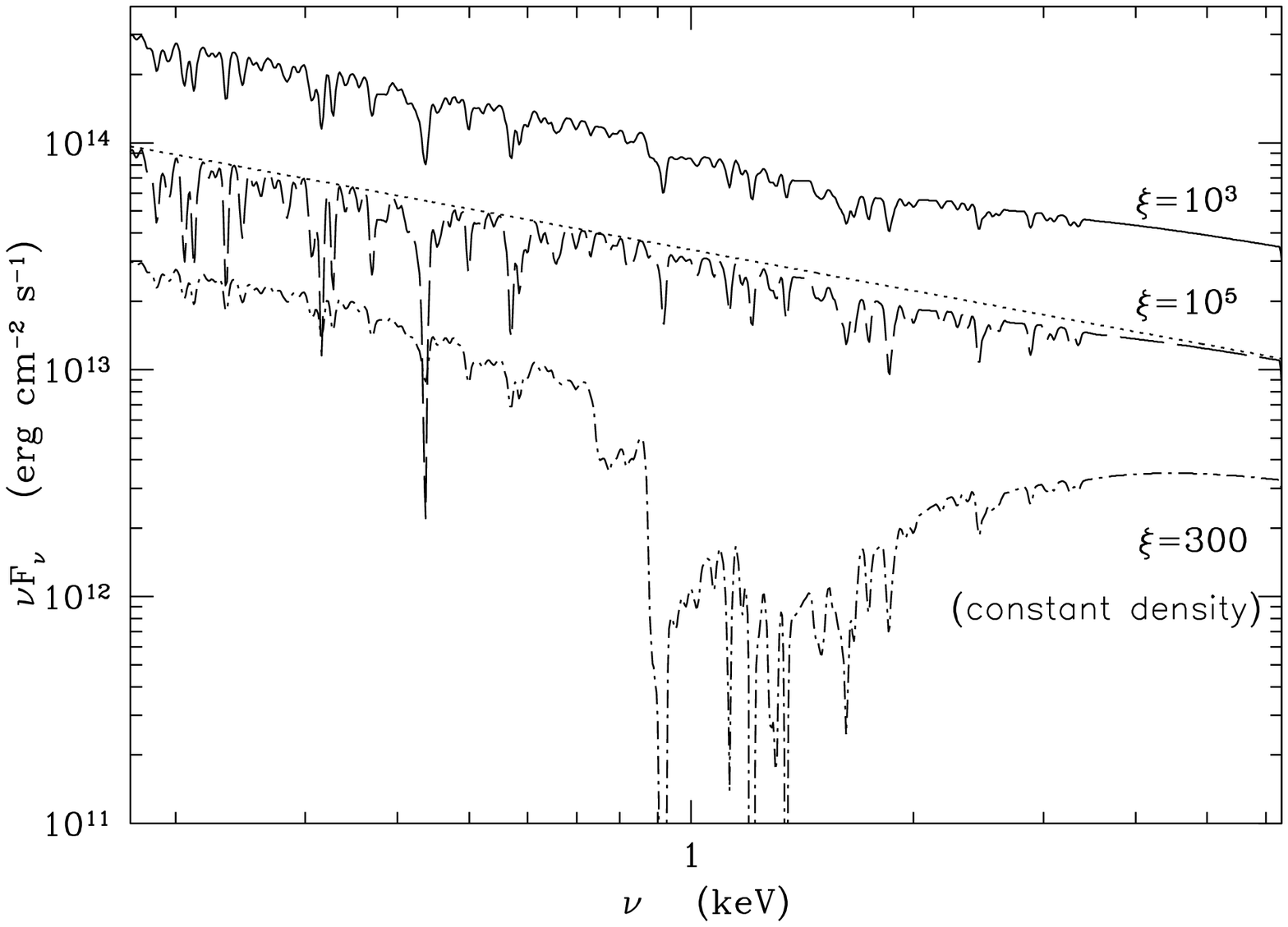}
  \caption{Spectral energy  distribution  
(spectral resolution 60) for models with  $\xi=10^5$
    and $N_{tot} =1.7 \times 10^{23}{\rm cm}^{-2} $ (dashed line),  
     $\xi=10^3$ and $N_{tot} =6.8 \times 10^{22}{\rm cm}^{-2} $ (solid line). 
  Constant density case for $\xi=300$ and $N_{tot} =1.7 \times 10^{23}{\rm cm}^{-2}$ is
 presented by dotted-dashed line.
  Both solid and dotted-dashed lines are arbitrarily moved  to see differences.
      Dotted line indicates the slope of incident continuum. }
 \label{fig:mod1}
\end{figure}

We have used code {\sc titan} (Dumont, Abrassart \& Collin 2000) to compute 
temperature and density structure of the warm absorber together with  
radiative transfer. The code is based on two-stream approximation
 and works in plane-parallel geometry. 
Radiative transfer is computed both in lines and
in continuum so the code can be used for very inhomogeneous even 
optically thick media. 
Thermal, ionization and statistical equilibrium of ions 
are computed in
complete non-LTE for 10 elements and all corresponding ions.
We consider following elemental abundances:
H: $1$, He: $0.085$, C: $3.3 \times 10^{-4}$, N: $9.1 \times 10^{-5}$,
O: $6.6 \times 10^{-4}$, Ne: $8.3 \times 10^{-5}$, Mg: $2.6 \times 
10^{-5}$,
Si: $3.3 \times 10^{-5}$, S: $1.6 \times 10^{-5}$, Fe: $3.2 \times 
10^{-5}$,.  We transfer totally 721 lines of different
ions, taking their oscillator strengths and energies from National 
Institute of Standards and Technology
(NIST, http://physics.nist.gov). Photo-electric absorption is fully 
taken into account.

In our models, we have assumed the shape of incident radiation   
consistent with this fitted to the data of Ton S180. 
We have adopted power law photon index $\Gamma=2.6$ (Sec~\ref{sec:cont}),
and spectral limits   
from $0.012$ up to $100$ keV as  we have concluded from broad band  
spectral energy distribution presented by Turner et al. (2002).

The assumption of constant pressure refers to the total: radiation and
gas pressure, and its value is self consistently determined by the code.
The ionization parameter on the top of the absorber $\xi_0$ and
the total column density are free parameters of our models. 
We have computed models for 
$\xi_0=10^5$ and $N_{tot}=1.7 \times 10^{23}{\rm cm}^{-2}$ named A1,
$\xi_0= 1000$ and $N_{tot}=6.8 \times 10^{22} {\rm cm}^{-2}$ named A2, and
$\xi_0=10^7$ and $N_{tot}=1.7 \times 10^{23}{\rm cm}^{-2}$ named A3.
Such values of column densities for given $\xi_0$ are  
upper limits above which optical depth of cloud
is large enough that 
strong absorption edges become present in outgoing spectra. 

Even for high $\xi_0$ , the maximum Compton temperature is low, 
$T_{Compt} = 2.1 \times 10^6$ K, because
the incident  spectrum is soft ($\Gamma =2.6$). 

In Fig.~\ref{fig:mod} we show the temperature structure 
for A1 (dashed line) (higher $\xi_0$ does not change this structure)
compared to A2 (solid line) and in Fig.~\ref{fig:mod1}
we show the corresponding spectra.
In case of higher $\xi_0$ (model A1) the 
temperature range 
within the medium is much broader 
and the effective spectrum from such region (presented in
Fig.~\ref{fig:mod1} dashed line) displays slightly stronger lines than for A2 
(solid line spectrum). 
We have enlarged range of frequencies to illustrate 
region observed in Ton S180 data. 

We have also  performed comparison of A1 model to the warm absorber with
constant density (Fig.~\ref{fig:mod} dotted-dashed line)
assuming that both clouds have the same total column densities 
and they are illuminated by the equivalent fluxes.
Thus,  choosing the volume density for constant density cloud,
the ionization parameter is determined by  the value of flux
the same as for stratified cloud.
We have chosen volume density equal $3 \times 10^{13} {\rm cm^{-3}}$, 
which implies $\xi_0=300$. 
Spectrum for constant density cloud presented in Fig.~\ref{fig:mod1} by
dotted-dashed line displays strong edges and 
cannot be applicable to explain our observations.
Decreasing volume density we increase ionization parameter,
and it may remove edges. We plan to make proper calculations 
in the future work to show how to distinguish between
low density constant cloud and stratified cloud.

We have also calculated theoretical values of the 
equivalent widths of transfered lines, the same
which are found in Ton S180 data. Results are presented
in Table~\ref{tab3}. 
EWs for high surface ionization (A1) are almost a factor of two 
higher that for  A2 model.
In our data we see slightly larger EWs and we cannot model them by 
increasing $\xi_0$,
since model A3 shows the same EWs as  A1.

Therefore, we have to introduce model labeled  A4 in Table~\ref{tab3}, which
has exactly the same parameters as A2 i.e. $\xi_0= 1000$ and 
$N_{tot}=6.8 \times 10^{22} {\rm cm}^{-2}$
and additionally we assume turbulent velocity $v_{turb}/c =3 \times 10^{-4}$.
Turbulent velocity does not change the total temperature structure, it only 
affects lines, making their EWs higher, as shown in  Table~\ref{tab3}.
For model A4 several lines, for instance 
Mg{\sc xi}, Si{\sc xiii}, S{\sc xii}
C{\sc vi},  and  Si{\sc xiv}) have EWs in quite good agreement with those 
achieved from observations (see Table~\ref{tab1}). 

In all presented models the ratio of 
$EW$[C{\sc vi}$_{(1-2)}]$ / $EW$[C{\sc vi}$_{(1-5)}]$ 
equals 0.7, in very good 
agreement with observed value of 0.6.

\begin{deluxetable}{llllll}
\renewcommand{\arraystretch}{1.5}
\tablenum{4}
\tablewidth{120 mm}
\tablecaption{Equivalent width [eVs] of absorption lines for several models computed 
by {\sc titan}
\label{tab3}}
\tablehead{
\colhead{Ion} &
\colhead{Transition} &
\colhead{A1} &
\colhead{A2} &
\colhead{A3} &
\colhead{A4}
}
\startdata

S{\sc viii} & 2s$^{2}$ 2p$^{5}$  $^{2}$P$^{0}$ - 2s$^{2}$ 2p$^{5}$ 3d  $^{2}$L
& 0.522 & 0.293 &0.506 &  1.431 \\

Fe{\sc xv} &  3s$^{2}$  $^{1}$S - 3s4p  $^{1}$P$^{0}$
& 0.406   &  0.191 & 0.392  & 1.164  \\


Si{\sc x} & 2s$^{2}$ 2p  $^{2}$P$^{0}$ - 2s$^{2}$ 3d  $^{2}$D
& 0.584 & 0.286 & 0.568 & 1.193  \\

S{\sc x} & 2s$^{2}$ 2p$^{3}$  $^{4}$S$^{0}$ -2s$^{2}$ 2p$^{2}$($^{3}$P) 3l  $^{4}$P 
& 0.585   &  0.305 & 0.567 & 1.549  \\


C{\sc v} & 1s$^{2}$  $^{1}$S - 2p  $^{1}$P$^{0}$ 
& 1.112  & 0.587 & 1.079 & 1.849   \\


S{\sc xi} & 2s$^{2}$ 2p$^{2}$  $^{3}$P - 2s$^{2}$ 2p 3d  $^{3}$L$^{0}$
& 0.680 & 0.333 & 0.659 & 1.500 \\

Mg{\sc xi} & 2p  $^{3}$P$^{0}$ - 4l  $^{3}$L
& 0.834 & 0.468 &0.809& 1.995 \\

Si{\sc xiii} & 2s  $^{1}$S - 3p  $^{1}$P$^{0}$ 
& 0.783  & 0.439 & 0.759 &  2.019   \\

S{\sc xii} & 2s$^{2}$ 2p  $^{2}$P$^{0}$ - 2s$^{2}$ 3d  $^{2}$D
& 0.719 & 0.348 & 0.697 & 1.580\\

C{\sc v} & 1s$^{2}$  $^{1}$S - 3p  $^{1}$P$^{0}$
& 1.287 & 0.708 & 1.249& 2.151 \\

C{\sc vi} & 1s  $^{2}$S - 2l  $^{2}$L 
& 1.186 &  0.599 & 1.136& 1.772 \\

C{\sc vi} & 1s  $^{2}$S - 5l  $^{2}$L
& 1.650  &  0.881 & 1.585& 2.585 \\

Fe{\sc xvii} & 2s$^{2}$ 2p$^{6}$  $^{1}$S - 2s$^{2}$ 2p$^{5} $3s  $^{1}$P$^{0}$ 
& 1.216  & 0.621 & 1.159 &  4.008  \\

Ne{\sc x} &  1s $^{2}$S - 3l $^{2}$L 
& 3.151   & 1.856 & 3.063 &  7.485  \\

Mg{\sc xii} & 1s $^{2}$S - 2l $^{2}$L
&  3.561 & 2.101 &  3.561 & 9.107\\

Mg{\sc xii} & 1s $^{2}$S - 4l $^{2}$L 
 & 5.224 & 2.947 &  5.330& 12.26\\

Si{\sc xiv} & 1s $^{2}$S - 2l $^{2}$L
& 4.594 & 2.426 &  4.598 & 11.56 \\

\enddata

\tablenotetext{a}{
l means that the configuration is a mixture of multiplets s p d}

\end{deluxetable}

\section{Broad band continuum} 
\label{sec:bbc}

In order to draw attention to the possible 
constraints resulting from detection of the broad disk lines
we model the broad band spectrum of Ton S180 assuming three basic
spectral components: (a) an illuminated accretion disk (b) an optically thick
Comptonizing plasma (c) an optically thin (non-thermal) Comptonizing plasma.
For low or moderate accretion rate we take 
the simplest Keplerian disk model, emitting locally as a black body
and parameterized fully by the black hole mass, $M$, and accretion rate, 
$\dot m$, given in Eddington units (taking efficiency 1/12 appropriate for 
Newtonian model). 
We consider also highly supercritical accretion disk applying the model
described by Kawaguchi (2002). This model includes the effect of
strong radial advection and the departure from the local black body
emission due to electron scattering which are essential at very high
accretion rates.

We assume that the energy generated in the inner disk 
illuminates outer disk, with the irradiating flux $\propto r^{-2-\beta}$, with 
$\beta = 0.1$. 
This irradiation effect may come from the scattering of
the disk radiation back toward the disk by the warm absorber as well as
from the direct interception of the radiation by the flaring outer disk.
We adopted a simple parameterization and did not model this irradiation
since the process is quite complex. It
would require assumption about the radial and angular distribution of the
density of the scattering medium (see e.g. Kurpiewski et al. 1997) and the
computations of the viscosity dependent disk vertical structure, as in
R\'o\.za\'nska et al. (1999).
We did not compute the disk vertical structure in order to
verify whether the outer  disk is flaring strongly enough to allow for such a
strong effect. 

We assume that the inner disk for ($ r < r_{skin}$) is
fully covered by the optically thick plasma. 
Thermal plasma parameters are: the optical
depth, $\tau$, and the temperature, $T$. 
Only soft photons from the inner disk come
 through this plasma, and we compute the spectrum 
using the simple code  based 
on Sunyaev \& Titarchuk (1980) description
(used for example by  Collin-Souffrin et al. 1996). We adopt the
plane parallel geometry and the generation of the photons at the bottom of the
zone. This component forms a soft X-ray component of the spectrum.
We next added a contribution from much hotter, or non-thermal, 
plasma component in order to provide
the appropriate hard X-ray tail but the parameters of this component
are poorly constraint and we do not address this
issue in detail.

Our model provides a one-parameter family of solutions for the broad band 
spectrum, i.e. we can find a good model for any mass of the black hole between
$1 \times 10^7 M_{\odot}$ and $1 \times 10^8 M_{\odot}$, but for a fixed mass
all other parameters ($\dot m$, $r_{skin}$, $\tau$, and $T$) are uniquely 
determined. The appropriate accretion rate is needed to reproduce the 
normalization of the spectrum in the optical band, where it is mostly given by
the product of $M^2 \dot m$ ( for illumination case). 
The value of $r_{skin}$ must be chosen in 
order not to overproduce/underproduce far-UV flux. The slope of the soft 
X-ray spectrum gives one constraint on the combination of $\tau$ and $T$, 
but for low values of $T$ which are always derived the normalization of 
the soft Comptonized spectrum also depends significantly on the plasma 
parameters (for a number of soft photons already specified by $r_{skin}$).

\begin{figure}[h]
  \plotone{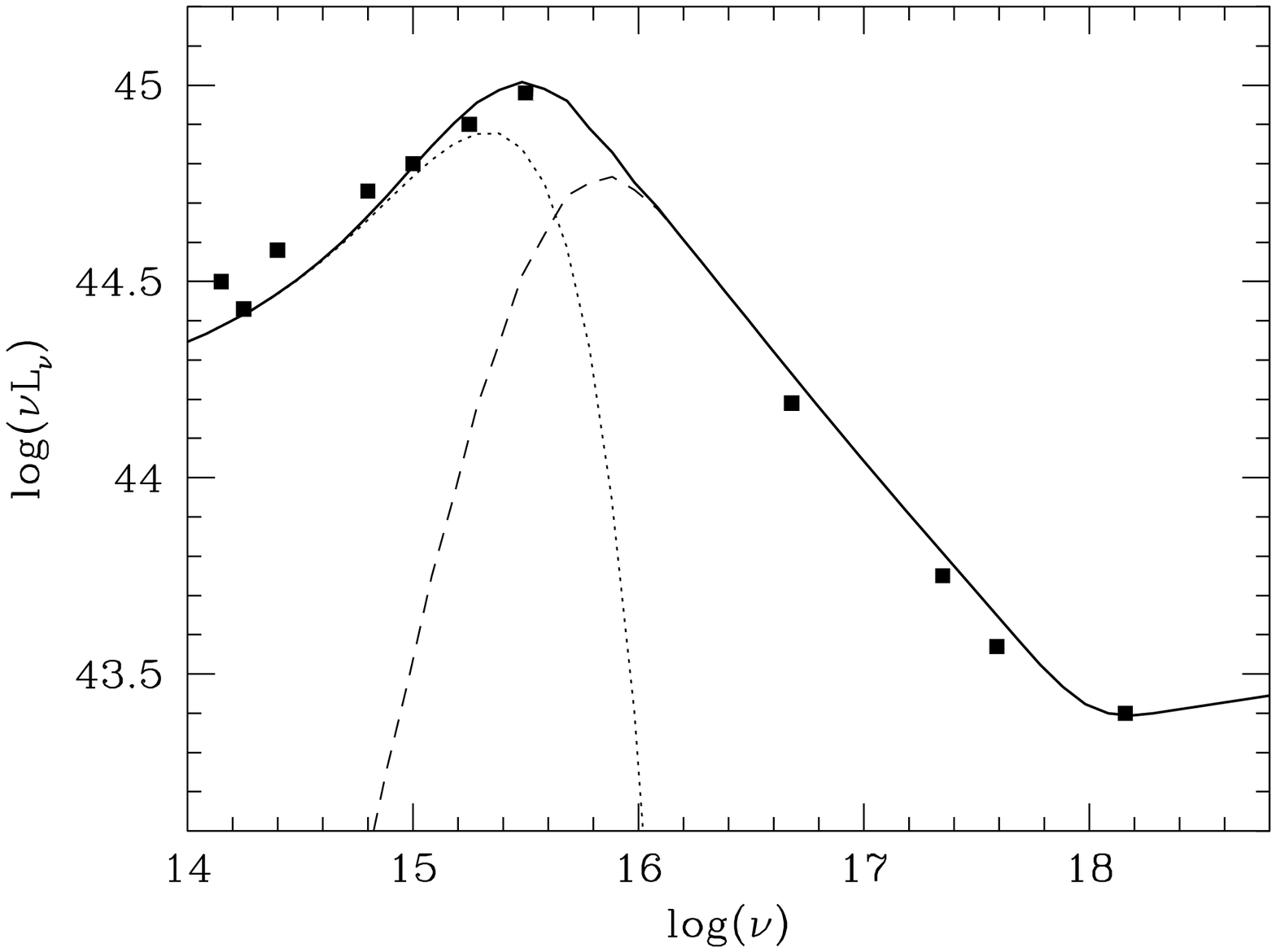}
  \caption{The model of the broad band spectrum of Ton S180 (solid line)
for the minimum acceptable accretion rate.
 Dotted line gives the spectrum of the outer uncovered part of the disk and 
dashed line shows the radiation from the inner disk after passing through 
the Comptonizing medium. Model parameters: $M = 10^8 M_{\odot}$, $\dot m = 0.07$, 
$r_{skin} = 20 R_g$,
$\tau = 17$, $T=0.65$ keV.}
 \label{fig:bbwidmo_a}
\end{figure}

\begin{figure}[h]
  \plotone{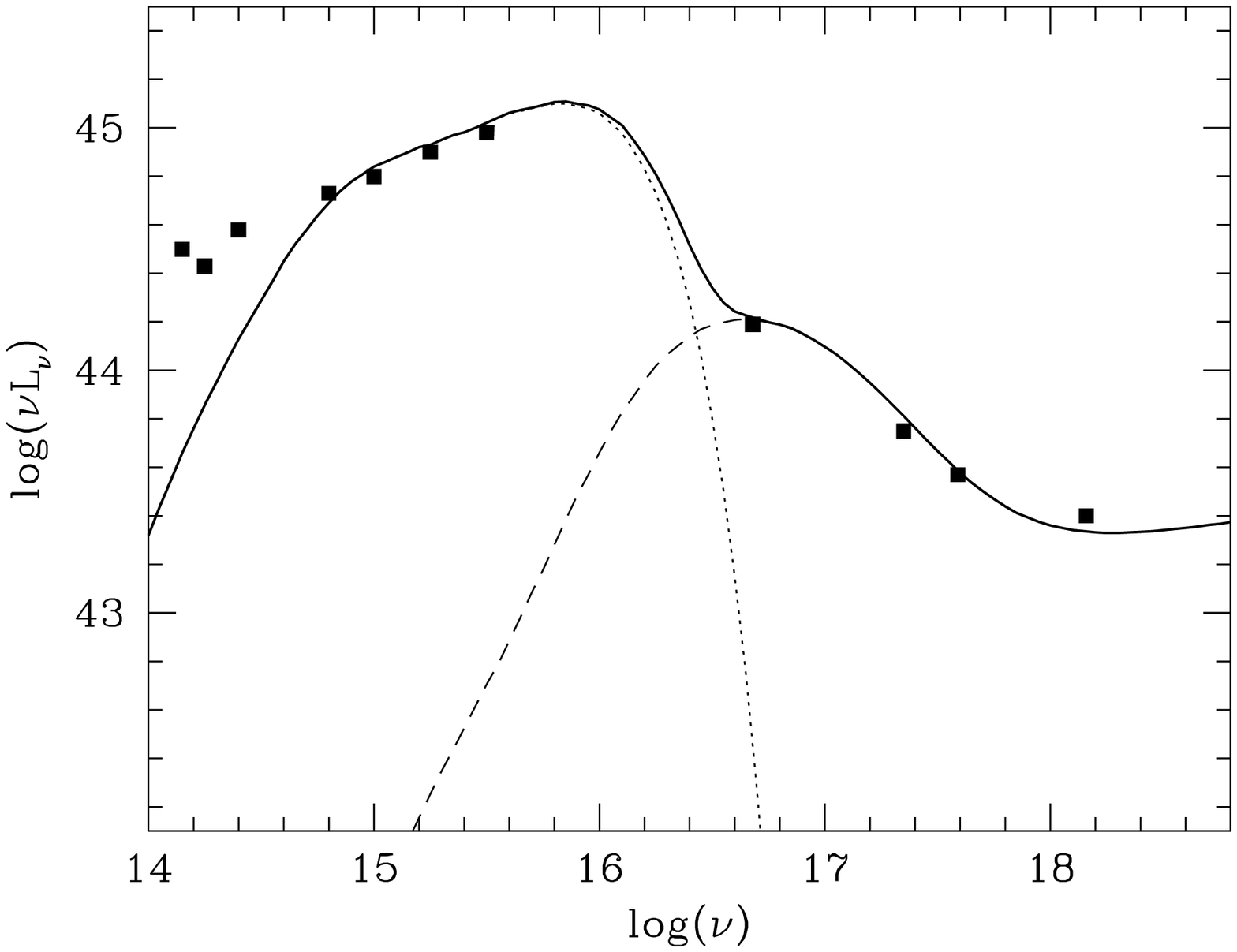}
  \caption{The model of the broad band spectrum of Ton S180 (solid line) for 
very high accretion rate (Kawaguchi 2002) .
 Dotted line gives the spectrum of the outer uncovered part of the disk and 
dashed line shows the radiation from the inner disk after passing through 
the Comptonizing medium. Model parameters: $M = 10^7 M_{\odot}$, $\dot m = 83$, 
$r_{skin} = 140 R_g$,
$\tau = 2.5$, $T=12$ keV.}
 \label{fig:toshi}
\end{figure}

Examples of the solution are therefore as follows:
for $M =  10^8 M_{\odot}$: we have $\dot m = 0.07$, $r_{skin} = 20 R_g$,
$\tau = 17$, $T=0.65$ keV, 
and for high accretion rate case:
$M =  10^7 M_{\odot}$: we have $\dot m = 83$, $r_{skin} = 140 R_g$,
$\tau = 2.5$, $T=12$ keV;  
The broad 
band data fit for the largest mass is shown in Fig.~\ref{fig:bbwidmo_a}
 and for the smallest mass --- in Fig.~\ref{fig:toshi}.

Those two extreme models differ considerably with the extension of the disk 
part covered with the hot skin. In the case of small mass, the disk 
model would overpredict the far-UV data points considerably if the large
portion of the inner disk is not fully 
covered with the thermal plasma. Plasma parameters are such that they make 
formation of any emission lines rather difficult.

In the case of high mass  
the inner disk does not really have to be covered,
as indicated by much smaller value of $r_{skin}$. Clumpy corona model
would be easily applicable to this case, or even the X-ray emission could 
come from some shocks in outflowing material at the rotation axis. The 
fraction of the disk covered may be always further reduced assuming that 
synchrotron emission of the plasma itself provides additional soft photons.
In such a case the lines from the irradiated disk are easily expected. 

\section{Discussion}
\label{sec:dis}

\subsection{Warm absorber}

Reanalyzing {\it Chandra} data of Narrow Line Seyfert 1 Galaxy - Ton S180
we were able to identify several narrow absorption lines due to the warm 
absorber. Such lines were neither discussed by Turner et al. (2001) and 
Turner et  al. (2002) where they presented the data for the first time 
nor seen in {\it XMM-Newton} data (Vaughna et al. 2002) because the lines
are indeed very faint. 
We detected lines of highly ionized carbon and other
elements, but we do not seem to detect any oxygen lines while in the FUSE
data (Turner et al. 2002) O{\sc vi} narrow absorption line 
components were seen. 

Five of those absorption lines detected in Ton S180 
(C{\sc vi}($E=0.3673$ keV), Ne{\sc x}, Mg{\sc xii}($E=1.471$ keV),
Mg{\sc xii}($E=1.840$ keV), and Si{\sc xiv})
are also seen in another Seyfert 1 galaxy
- NGC 3783 analyzed by  Kaspi et al. (2002), with similar values of EWs
(0.46, 2.3, 4.3, 1.6 and 6.6 eV respectively).   
Also Kaastra et al. (2002) in NGC 5548 have found the same C{\sc v},
C{\sc vi}($E=0.3673$ keV), Ne{\sc x}, Mg{\sc xii}($E=1.471$ keV), and
Si{\sc xiv}, with following EWs - 0.26, 0.99, 0.47, 3.09, and 3.88 eV.

We found a lower limit for the warm absorber column of
$ 5.1 \times 10^{21} $ cm$^{-2}$ from the equivalent with of the one of
C {\sc vi} lines (see Sect.~\ref{sec:abs}) and an upper limit of 
$1 \times 1.7 \times 10^{23} $ cm$^{-2}$ from our model of the warm absorber (see
Sect.~\ref{sec:mod}). The upper limit is not firm since it is sensitive
to the adopted turbulent velocity as well as abundances. 

\subsection{Broad-band continuum}

The {\it Chandra} data show that the effect of the warm absorber 
does not modify 
significantly the X-ray 
spectrum in the case of Ton S180. Therefore our analysis do not change the
overall picture of the Ton S180 continuum, as shown by Turner et al. (2002)
and discussed by Vaughan et al. (2002). In particular, the huge soft X-ray
excess is reasonably explained only within the frame of the Comptonization 
model. 

Variability arguments presented by Vaughan et al. (2002) in favor of hybrid
plasma model (thermal plasma being responsible for the soft X-ray slope and 
non-thermal plasma component leading to hard X-ray power law) are convincing.
However, the data do not allow to determine all the hybrid plasma parameters
uniquely, as explained by Vaughan et al. (2002), so in the present paper 
we have used much simpler model in order to draw attention to the possible 
constraints resulting from detection of the broad disk lines.

Those emission lines can be present in the broad band spectrum rather
for higher black hole mass $M=10^8 M_{\odot}$ and accretion rate $\dot
m =0.07 $, since with those assumptions hot skin does not cover disk
completely.  This conclusion is opposite to the black hole mass
estimations done for Ton S180 by Wang \& Lu (2001), where $M=1.3
\times 10^7 M_{\odot}$.  But our result is consistent with studies done
by Siemiginowska (1997), where simple disk corona model was fitted to the
{\it Beppo SAX} data (those presented by  Comastri et al. 1998). 

\section{Conclusions}
\label{sec:conc}

We argue, contrary to the previous papers by Turner et al. (2001) and
Vaughan et al. (2002) that spectral features in absorption and even
in emission are present in the {\it Chandra} data of Ton S180.
The quality of the data does not allow us to determine column densities of 
individual metals, we can only put lower limit on column density of C{\sc vi}.

The lack of edges in the spectrum gives upper limit on total column density,
which is $1.7 \times 10^{23}$ cm{$^{-2}$} for the best computed model 
($\xi_0=10^5$).

A few emission lines fitted to the Ton S180 {\it Chandra} data have
lower equivalent widths by factor of 6  than those 
reported by Mason et al. (2002) for Seyfert I galaxy Mrk 766, but
of similar order as disk lines found by Kaastra et al. (2002) in NGC 5548.

Estimation of the mass of the black hole in Ton S180 based on FWHM of the
$H_{\beta}$ line and the formula of Kaspi et al. (2000) for the relation 
between the size of the Broad Line Region and the object luminosity at 
5100 \AA~ rather suggests small black hole mass, 
$M = 1.3 \times 10^7 M_{\odot}$ 
(Wang \& Lu 2001),
in contradiction with our conclusion based on the detection of the
soft X-ray lines.

\vspace{0.3cm}

{\sl Acknowledgments.}  We are grateful to Suzy Collin for discussion
regarding spectral models of warm absorber computed by {\sc titan}.
We thank Liz Galle for help in processing the archival data and
Antonella Fruscione and Fabrizio Nicastro for discussion.  This work was
supported in part by grant 2P03D00322 of the Polish State Committee
for Scientific Research and by Jumelage/CNRS No. 16 ``Astronomie
France/Pologne''. AS acknowledge support through NASA contract
NAS8-39073(CXC)


\clearpage
\end{document}